\documentclass[12pt]{article}
\parskip=\medskipamount                 
\thicklines                     
\usepackage{amsfonts,amssymb,amsmath,amscd}
\usepackage{bbm,MnSymbol}
\usepackage{graphicx, epstopdf}
\usepackage{euscript,eufrak}
\usepackage[ps,dvips,matrix,arrow,frame,import,curve,color]{xy}

\newcommand{\bul}{\bullet}
\newcommand{\Tr}{{\rm Tr}}

\newcommand{\ZZ}{{\mathbb Z}}
\newcommand{\RR}{{\mathbb R}}
\newcommand{\CC}{{\mathbb C}}

\renewcommand{\d}{{\rm d}}

\newcommand{\tpsi}{{\tilde\psi}}
\newcommand{\teta}{{\tilde\eta}}
\renewcommand{\SS}{{\rm S}}

\newcommand{\cO}{{\mathcal O}}
\newcommand{\bpartial}{{\bar\partial}}

\newcommand{\bz}{{\bar z}}

\newcommand{\cA}{{\mathcal A}}

\newcommand{\Hom}{{\rm Hom}}

\newcommand{\Z}{\mathcal Y}

\newcommand{\cM}{{\mathcal M}}

\newcommand{\cN}{{\mathcal N}}
\newcommand{\End}{{\rm End}}
\newcommand{\cW}{{\mathcal W}}

\newcommand{\Sym}{{\rm Sym}}

\newcommand{\bI}{{\bar I}}
\newcommand{\bJ}{{\bar J}}

\newcommand{\cF}{{\mathcal F}}

\newcommand{\frg}{{\mathfrak g}}

\newcommand{\LG}{{{}^L G}}
\newcommand{\ra}{\rightarrow}

\newcommand{\trho}{{\tilde\rho}}

\newcommand{\tchi}{{\tilde\chi}}

\newcommand{\mT}{{\mathsf T}}
\newcommand{\frt}{{\mathfrak t}}
\newcommand{\coch}{{\rm cochar}}
\newcommand{\mTL}{{{}^L\mathsf T}}
\newcommand{\bsigma}{{\bar\sigma}}
\renewcommand{\P}{{\mathcal P}}

\newcommand{\Ad}{{\rm Ad}}
\newcommand{\Mor}{{\rm Mor}}
\newcommand{\Ob}{{\rm Ob}}
\newcommand{\dV}{{\mathbb V}}
\newcommand{\A}{{\mathsf A}}
\newcommand{\B}{{\mathsf B}}
\newcommand{\C}{{\mathsf C}}
\newcommand{\D}{{\mathsf D}}
\newcommand{\X}{{\mathbb X}}
\newcommand{\Y}{{\mathbb Y}}
\renewcommand{\Z}{{\mathbb Z}}
\renewcommand{\1}{{\mathbf 1}}

\newcommand{\KK}{{\mathfrak K}}
\newcommand{\LL}{{\mathfrak L}}
\newcommand{\cL}{{\mathcal L}}
\newcommand{\cC}{{\mathcal C}}
\newcommand{\frA}{{\mathfrak A}}
\newcommand{\dW}{{\mathbb W}}
\newcommand{\cP}{{\mathcal P}}

\newcommand{\cD}{{\mathcal D}}
\newcommand{\Obul}{{\Omega^{0,\bul}}}
\newcommand{\btau}{{\bar\tau}}
\newcommand{\bphi}{{\bar\phi}}
\newcommand{\stard}{\medstar}

\begin{document}

\title{Surface operators in four-dimensional topological gauge theory and Langlands duality}

\author{Anton Kapustin, Kevin Setter, Ketan Vyas \\ {\it \small California
Institute of Technology}}

\begin{titlepage}
\maketitle

\abstract{We study surface and line operators in the GL-twisted $N=4$ gauge theory in four dimensions. Their properties depend on the parameter $t$ which determines the BRST operator of theory. For $t=i$ we propose a complete description of the 2-category of surface operators in terms of module categories. We also  determine the monoidal category of line operators which includes Wilson lines as special objects. For $t=1$ and $t=0$ we only discuss surface and line operators in the abelian case.  Applications to the categorification of the local geometric Langlands duality and its quantum version are briefly described. In the appendices we discuss  several 3d and 2d topological field theories with gauge fields. In particular, we explain a relationship between the category of branes in the gauged B-model and the equivariant derived category of coherent 
sheaves.}

\end{titlepage}

\section{Introduction}

Recently geometric Langlands duality has been interpreted \cite{KW} as a consequence of an isomorphism between two different topological gauge theories in four dimensions whose gauge groups are related by Goddard-Nuyts-Olive duality \cite{GNO}. These topological gauge theories are obtained by twisting $N=4$ $d=4$ super-Yang-Mills theory by means of the so called GL-twist. The goal of this paper is to classify and study certain nonlocal observables (surface operators) in GL-twisted gauge theory, with a view towards strengthening and extending the geometric Langlands duality. 

From a mathematical point of view, a 4d Topological Field Theory (TFT) is a gadget which assigns a number to a compact oriented four-manifold, a vector space to a compact oriented three-manifold, a category to a Riemann surface, a 2-category to a circle, and a 3-category to a point. These data have varying amount of extra structure whose complexity is inversely related to the categorical level. For example, the category attached to a Riemann surface is acted upon by the mapping class group of the surface. The 2-category attached to a circle is a braided monoidal 2-category. The 3-category attached to a point is self-dual. 

From a physical viewpoint, the category attached to a Riemann surface $\Sigma$  can be interpreted as the category of boundary conditions for an effective 2d TFT obtained by compactifying the 4d TFT on $\Sigma$. The 2-category attached to a circle is the 2-category of boundary conditions for the 3d TFT obtained by compactifying the 4d TFT on the circle. Alternatively, it is the 2-category of surface operators in the 4d TFT. 

Equivalences of 4d TFTs therefore have implications for TFTs in lower dimensions. This was exploited in \cite{KW} where Montonen-Olive duality was shown to imply an equivalence of categories of boundary conditions for certain topological sigma-models in 2d. Going one dimension higher, we may consider GL-twisted $N=4$ gauge theories with gauge groups $G$ and $\LG$ compactified on a circle. The corresponding 3d TFTs are also isomorphic, so their 2-categories of boundary conditions must be equivalent. From the 4d point of view, the 2-category of boundary conditions in the 3d TFT is simply the 2-category of surface operators, where we ``forgot'' about the braided monoidal structure. 

It would be a mistake to think that the independence of the GL-twisted gauge theory on the metric means that we can let the circle radius to be zero and interpret the 3d TFT as a twist of $N=8$ $d=3$ super-Yang-Mills theory. This would imply that certain topological versions of $N=8$ $d=3$  SYM theories with gauge groups $G$ and $\LG$ are equivalent. However, this is not correct, as can be seen already in the abelian case. Indeed, compactifying $U(1)$ gauge theory on a circle of finite radius gives rise to a periodic scalar in the effective 3d theory (the holonomy of the gauge field along the compact direction). The period becomes very large as the circle radius goes to zero. But no matter how large the period is, the theory of a periodic scalar is different from that of an $\RR$-valued scalar. For example, the former theory admits disorder loop operators where the scalar has a nonzero winding, while the latter theory does not. These disorder loop operators are crucial for maintaining the Montonen-Olive duality, as we will see below.

In this paper we study the aspects of Montonen-Olive duality which can be understood in 3d terms. Namely, we study the 2-category of surface operators in the GL-twisted 4d theory, regarding it as a 2-category of boundary conditions in a 3d TFT. Some examples of surface operators in 4d TFT have been discussed in \cite{GW}, but we will see that there are much more general ones.

The GL-twisted theory depends on a complex parameter $t$ which determines the BRST operator of the theory. For $t=i$ the analysis of surface operators is relatively simple, since the 3d theory turns out to be of a rather familiar kind (the gauged Rozansky-Witten model). In this case we propose a description of the full 2-category of surface operators. For $t=1$ and $t=0$ the 3d theory is rather unusual, and in this paper we analyze it only in the abelian case. We also describe how abelian electric-magnetic duality acts on surface operators and line operators on them. Our description of the 2-categories of surface operators at $t=i$ and $t=1$ can be combined with electric-magnetic duality to give a statement that certain 2-categories attached to the group $U(1)$ are equivalent. This can be viewed as a 2-categorical analogue of the results of \cite{KW}.

GL-twisted gauge theory at $t=0$ is particularly interesting. This value of $t$ is preserved by Montonen-Olive duality and, as explained in \cite{Kap:qGL}, the corresponding 4d TFT provides a natural setting for understanding Quantum Geometric Langlands Duality.  On the other hand, it was shown in \cite{Kap:qGL} that this TFT does not admit either 't Hooft or Wilson line operators. This presents a problem for the mathematical formulation of the quantum Langlands duality. We will see that while the category of bulk line operators at $t=0$ is indeed rather boring, the abelian theory admits surface operators (not of Gukov-Witten type) whose categories of line operators are quite rich and are acted upon in a nontrivial way by electric-magnetic duality. Hopefully, these observations can be extended to the nonabelian case.

Reduction to 3d is useful only insofar as one can understand and classify boundary conditions in 3d TFTs. Up to now, the only 3d TFT where this has been achieved has been the Rozansky-Witten model \cite{KRS,KR}. In this paper we encounter a number of rather unfamiliar 3d TFTs, such a B-type topological gauge theory in 3d and a gauged version of the Rozansky-Witten model.  These theories are of independent interest, and we describe some of their properties, including nonlocal observables, in appendices C and D. We will also need to understand categories of branes in certain 2d TFTs obtained by reducing 3d TFTs on an interval. Sometimes these 2d TFTs are topological sigma-models (of type A or B), and then the categories of branes are known (the Fukaya-Floer category and the derived category of coherent sheaves, respectively).  In other cases they are gauged topological sigma-models which again can be of type A or B. In appendix A we describe the category of branes for the gauged B-model whose target is a point (i.e. for the B-type gauge theory). In appendix B we describe the category of branes for the gauged B-model with a general target space. Not surprisingly, we find that the category of branes in this TFT is closely related to the equivariant derived category of coherent sheaves and under certain assumptions is equivalent to it. In appendix F we study yet another 2d TFT, the A-type topological gauge theory. We show that when the gauge group is $U(1)$, this model is isomorphic to a B-model whose target is a graded bosonic manifold. This isomorphism allows one to identify the category of branes in the A-type gauge theory  This result may be regarded as a physical counterpart of the equivalence between the $U(1)$-equivariant constructible derived category of sheaves over a point and the derived category of modules over the cohomology of the classifying space of $U(1)$ \cite{BL}. 

A.K. would like to thank R.~ Bezrukavnikov, A.~Braverman, D. ~Orlov, V.~Lunts and especially L.~Rozansky for useful discussions.  K.S. acknowledges the support of the Jack Kent Cooke Foundation. This work was supported in part by the DOE grant DE-FG02-92ER40701.

\section{Topological field theory, categories and 2-categories: a brief summary}

This section is devoted to a brief review of the relationship between 2-categories and Topological Field Theory in two, three, and four dimensions. Readers who are familiar with this subject may skip the section.

\subsection{Two-dimensional TFT and categories of branes}

For us, a category is a generalization of an algebra, "an algebra with many objects". That is, instead of one vector space $V$ with a multiplication map $V\otimes V\ra V$ we have a set $\Ob$, a collection of vector spaces $V_{\A\B}$, $\A,\B\in \Ob$, and composition maps $V_{\A\B}\otimes V_{\B\C}\ra V_{\A\C}$. These composition maps must be associative, in an obvious sense. In particular, for each $\A$ the space $V_{\A\A}$ is a (possibly noncommutative) algebra. We will assume in addition that all these algebras have unit elements. 

The set $\Ob$ is called the set of objects, and the vector spaces $V_{\A\B}$ are called spaces of morphisms. An element of $V_{\A\A}$ is called an endomorphism of $\A$, and $V_{\A\A}$ is called the endomorphism algebra of $A$. It is common to denote $V_{\A\B}=\Mor(\A,\B)$ and $V_{\A\A}=\Mor(\A,\A)=\End(\A)$. In physical applications the vector spaces are always complex and usually have integral grading (by some sort of $U(1)$ charge). 

It is well-known by now that the set of boundary conditions in a 2d TFT has the structure of a category. The set $\Ob$ of this category is the set of boundary conditions, and the vector space $V_{\A\B}$ is the space of states of the TFT on an (oriented) interval with boundary conditions $\A$ and $\B$. Composition of morphisms arises from the fact that the space $V_{\A\B}$ can be interpreted as the space of local operators sitting at the junction of two segments of the boundary with boundary conditions $\A$ and $\B$ (Fig.1), and from the fact that local operators can  be fused together (Fig.2).
\begin{figure}[htbp]  \label{fig:SOGL_op}
\centering
\includegraphics[height=2in]{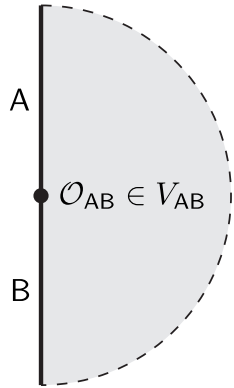}
\caption{Morphisms in the category of boundary conditions correspond to local operators sitting at the junction of two segments of the boundary.}
\end{figure}
\begin{figure}[htbp] \label{fig:SOGL_opfusion}
\centering
\includegraphics[height=2in]{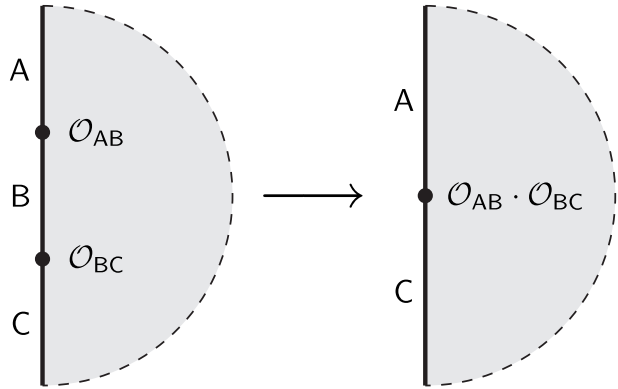}
\caption{Composition of morphisms is achieved by merging the insertion points of the local operators.  We use  $\cdot$ to denote this operation.}
\end{figure}

In the mathematical literature it is common to denote objects by marked points and elements of vector spaces $V_{\A\B}$ by arrows connecting the points. From the physical viewpoint it is more natural to denote objects by marked segments of an oriented line, and morphisms by points sitting at the junction of two consecutive segments. 

Let us recall two simple examples of 2d TFTs which will be important for us. The first one is a B-model with a target $X$, where $X$ is a complex manifold with a holomorphic volume form (i.e. a possibly noncompact Calabi-Yau manifold). The corresponding category of boundary conditions has been argued to be equivalent to the bounded derived category of coherent sheaves on $X$, which is denoted $D^b(Coh(X))$. Its objects can be thought of as complexes of holomorphic vector bundles on $X$ or  complex submanifolds of $X$. The second one is an A-model with target $Y$, where $Y$ is a symplectic manifold. The corresponding category of boundary conditions is believed to be equivalent to a version of the Fukaya-Floer category. Its simplest objects are Lagrangian submanifolds of $Y$ equipped with unitary vector bundles with flat connections. 
In the 2d context one usually refers to boundary conditions as branes and talks about A-branes and B-branes. 

A and B-models do not exhaust the possibilities even in two dimensions. Below we will encounter other, less familiar, 2d TFTs and their categories of branes.

\subsection{Two-dimensional TFTs and 2-categories}

A boundary of a 2d TFT can be regarded as a boundary between a nontrivial TFT and a trivial TFT. More generally, one can consider boundaries between arbitrary pairs of 2d TFTs. Such boundaries may be called defect lines, or walls. The set of all walls between a fixed pair of TFTs has the structure of a category. To see this, let $\X$ and $\Y$ denote our chosen pair of TFTs, and let $\bar\X$ denote the theory $\X$ with a reversed parity. By folding back the worldsheet at the wall location (see Fig. 3), we see that a wall between $\X$ and $\Y$ is the same as a boundary of the theory $\bar\X\times\Y$. 
\begin{figure}[htbp]  \label{fig:SOGL_foldingtrick}
\centering
\includegraphics[height=2in]{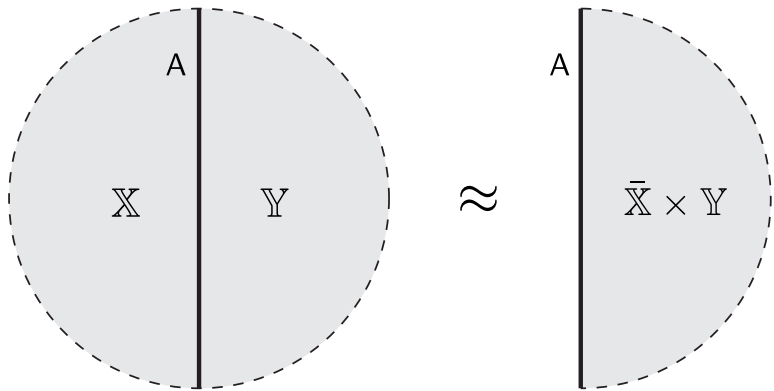}
\caption{A wall separating theories $\X$ and $\Y$ is equivalent to a boundary of theory $\bar\X\times\Y$.}
\end{figure}
Thus we may appeal to the previous discussion and conclude that walls are objects of a category $\dV_{\X\Y}$. Given any two walls $\A,\B\in\Ob(\dV_{\X\Y})$, the space of morphisms from $\A$ to $\B$ is the space of local operators which can be inserted at the junction of $\A$ and $\B$. Composition of morphisms is obtained by fusing local operators sitting on a defect line. 

There is an obvious ``fusion'' operation on the set of walls: given a wall between theories $\X$ and $\Y$ and a wall between theories $\Y$ and $\Z$ we may fuse them and get a wall between theories $\X$ and $\Z$ (Fig. 4).   One can describe the situation mathematically by saying that the set of 2d TFTs has the structure of a 2-category. A 2-category has objects, morphisms, and 2-morphisms (morphisms between morphisms). In the present case, objects are 2d TFTs. The set of morphisms from an object $\X$ to an object $\Y$ is the set of walls between theories $\X$ and $\Y$. Fusion of walls gives rise to a way of composing morphisms. Given any two walls between the same pair of TFTs, the space of 2-morphisms between them is the space of local operators which can be inserted at the junction of these two walls. 

One can put this slightly differently and say that a 2-category has a collection of objects (which are 2d TFTs in our case), and for any pair of objects $\X$ and $\Y$ one has a category of morphisms $\dV_{\X\Y}$ (which is the category of walls in our case). Fusion of walls means that there is a way to ``compose'' categories of morphisms. That is, given an object $\A$ of the category $\dV_{\X\Y}$ and an object $\B$ of the category $\dV_{\Y\Z}$ there is a rule which determines an object $\A\otimes \B$ of the category $\dV_{\X\Z}$. 
\begin{figure}[htbp]  \label{fig:SOGL_wallfusion}
\centering
\includegraphics[height=2in]{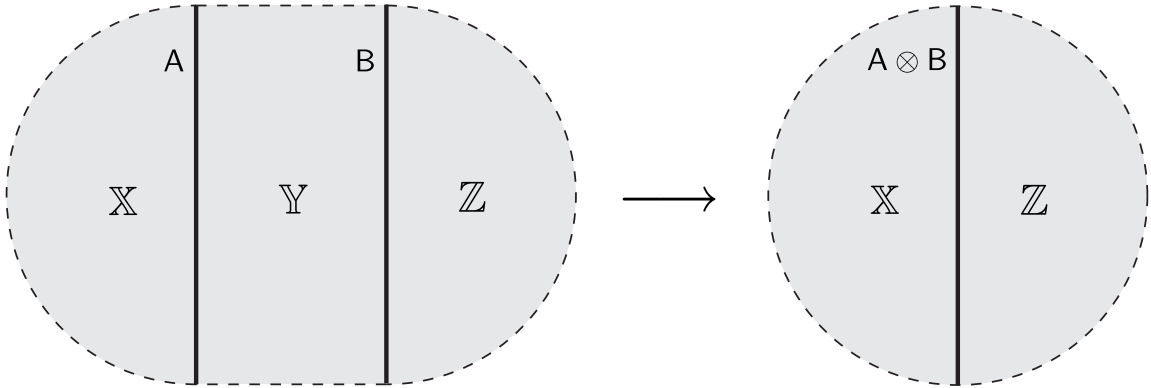}
\caption{1-morphisms of the 2-category of 2d TFTs correspond to walls, and composition of 1-morphisms corresponds to fusing  walls. This operation is denoted $\otimes$.}
\end{figure}
This is not all though: one can fuse not only walls, but walls with local operators inserted on them (Fig. 5). This determines composition maps on 2-morphisms.
\begin{figure}[htbp]  \label{fig:SOGL_opwall}
\centering
\includegraphics[height=2in]{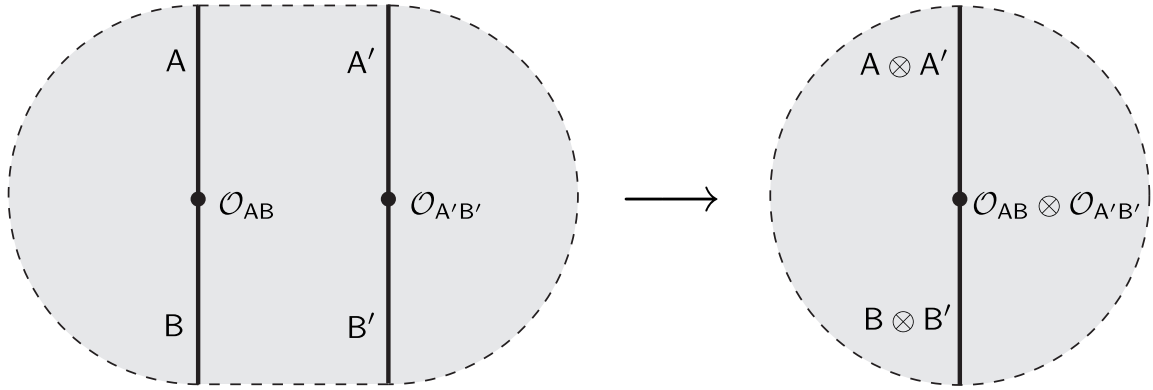}
\caption{Composition of 2-morphisms of the 2-category of 2d TFTs is achieved by fusing the walls on which they are inserted.  The corresponding operation is denoted $\otimes$.}
\end{figure}
This new composition is different from the composition of local operators regarded as morphisms in the category $\dV_{\X\Y}$. The composition maps enjoy various properties which can be deduced by staring at the pictures of fusing walls and making use of the metric independence. For example, the old and new compositions commute, as illustrated in Fig. 6.
\begin{figure}[htbp]  \label{fig:SOGL_commutative}
\centering
\includegraphics[height=3.5in]{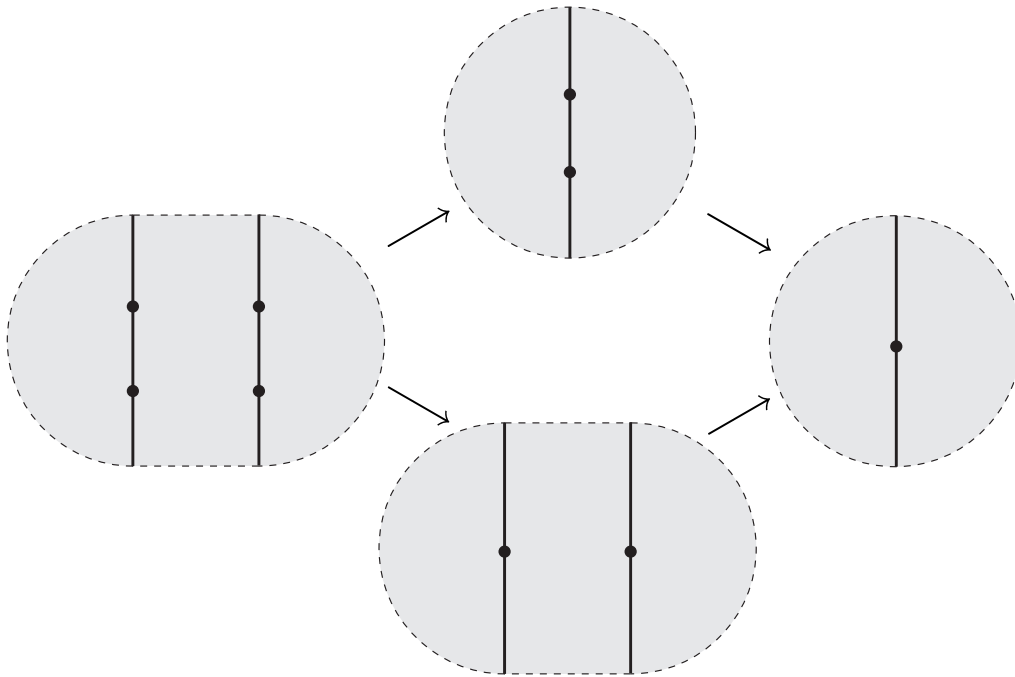}
\caption{Local operators inserted on walls may be regarded either as 2-morphisms of the category of 2d TFTs, in which case composition corresponds to fusing ``horizontally," or they may be regarded as morphisms of the category of boundary conditions, in which case composition corresponds to fusing ``vertically".  These two operations commute. }
\end{figure}

Even if one is interested in a particular TFT, the notion of a 2-category is useful. Namely, defect lines in a 2d TFT form a 2-category with a single object. In this case there is only one category of morphisms, $\dV_{\X\X}$, with additional structure coming from the fact that fusing two defect lines gives another defect line in the same theory. This structure allows one to define a rule for ``tensoring'' objects of $\dV_{\X\X}$:
$$
(\A,\B)\mapsto \A\otimes \B\in \Ob(\dV_{\X\X})
$$
and morphisms
$$
\Mor(\A,\B)\otimes \Mor(\C,\D)\ra \Mor(\A\otimes \C,\B\otimes \D)
$$
This tensoring does not need to be commutative, in general. A category with such additional ``tensor'' structure is called a monoidal category. It should be clear from the above that a monoidal category is the same thing as a 2-category with a single object.

Among all defect lines in a given 2d TFT there is a trivial defect line $\1$ which is equivalent to no defect at all. We may call it the invisible defect line. It is the identity object in the monoidal category of defect lines, in the sense that fusing it with any other defect line $\A$ gives back $\A$. Endomorphisms of the trivial defect line (i.e. elements of the vector space $\Mor(\1,\1)$) are the same as local operators in the bulk.

The simplest example of a monoidal category is the category of vector spaces, with the usual tensor product. It can be regarded as the 2-category of defect lines in a trivial 2d TFT (say, a topological sigma-model whose target is a point). Defect lines in Landau-Ginzburg TFTs have been studied in \cite{LGdefects1, LGdefects2}. 

One can fuse a defect line in a given 2d TFT with any boundary condition and get a new boundary condition in the same TFT. This defines an ``action'' of the monoidal category of defect lines on the category of branes. Mathematically this can be described using the notion of a module category. Since a monoidal cateory is a categorification of the notion of an algebra, it is natural to define a module category over a monoidal category as a categorification of the notion of a module over an algebra. A definition of a module category $\dW$ over a monoidal category $\dV$ involves a rule for ``multiplying'' an object on $\dW$ by an object of $\dV$:
$$
(\A,\C)\mapsto \A\cdot \C\in\Ob(\dW),\quad \forall \A\in \Ob(\dW),\ \forall \C\in\Ob(\dV),
$$
as well as a rule for multiplying morphisms, i.e. a map
$$
\Mor_\dW(\A,\B)\otimes \Mor_\dV(\C,\D)\ra \Mor_\dW (\A\cdot \C,\B\cdot\D).
$$
The latter rule encodes the fact that we can fuse a junction of two defect lines with a junction of two boundary conditions and get a new junction of two new boundary conditions. 

An important idea which we systematically use in this paper is that some properties of codimension-2 defects can be studied using dimensional reduction. We have already seen a simple example of this: the space of local operators sitting at the junction of two boundary conditions $\A$ and $\B$ can be thought of as the space of states of a 1d field theory (i.e. quantum mechanics) obtained by compactifying the 2d TFT on an interval with boundary conditions $\A$ and $\B$. Another example is the space of local operators in the bulk. It is well known that it can be identified with the space of states of a 1d field theory obtained compactifying the 2d TFT on a circle. The argument is essentially the same in both cases. After one excises a tubular neigborhood of the local operator, the operator insertion is replaced by a boundary whose collar neighborhood looks like $\RR_+\times I$ in the first case and $\RR_+\times S^1$ in the second case. Then one uses the fact that in a TFT the size of the tubular neighborhood does not matter, and one can regard any boundary condition on the newly created boundary as a local operator. 

It is important to note that this reinterpretation of codimension-2 defects in terms of a lower-dimensional theory causes ``information loss''. For example, we cannot compute the composition $V_{\A\B}\otimes V_{\B\C}\ra V_{\A\C}$ if we view the vector spaces involved as spaces of states of three 1d field theories. Similarly, we cannot see the commutative algebra structure on the space of bulk local operators if we view it as the space of states of a 1d field theory.

\subsection{Three-dimensional TFT and 2-categories of boundary conditions}

When we move to dimension three, we find that boundary conditions in a 3d TFT also form a 2-category. To see this, let us first consider a trivial 3d TFT (say, a 3d topological sigma-model whose target is a point). Even though there are no bulk degrees of freedom in this case, we may consider putting any 2d TFT on the boundary. Thus the set of all boundary conditions for the trivial 3d TFT is the set of all 2d TFTs, which form a 2-category. Walls between 2d TFTs now can be interpreted as defect lines on the 2d boundary of a 3d worldvolume. The monoidal category of defect lines for a given 2d TFT can be reinterpreted as the monoidal category of boundary line operators for a particular boundary condition.

If the 3d TFT in the bulk is nontrivial, we can still couple it to a 2d TFT on the boundary.  Different boundary conditions are distinguished by the type of 2d TFT on the boundary and by its coupling to the bulk degrees of freedom. For a concrete example of how this works in the Rozansky-Witten 3d TFT, see \cite{KRS}. Again one may consider boundary defect lines separating different boundary conditions, and their fusion and fusion of local operators on boundary defect lines can be described by a 2-category structure on the set of boundary conditions. If we focus on a particular boundary condition, then the set of defect lines on this boundary has the structure of a monoidal category.

Mimicking what we did in 2d, we may consider walls, or surface operators, between different 3d TFTs. The set of walls between any two 3d TFTs $\KK$ and $\LL$ has the structure of a 2-category. One way to see it is to fold the worldvolume along the defect surface and reinterpret the wall as a boundary condition for the theory $\bar\KK\times\LL$, where $\bar\KK$ is the parity-reversal of the theory $\KK$.  Furthermore, just like in 2d, we can fuse walls with defect lines and local operators on them (Fig. 7).
\begin{figure}[htbp] \label{fig:SOGL_surfacefusion}
\centering
\includegraphics[height=1.8in]{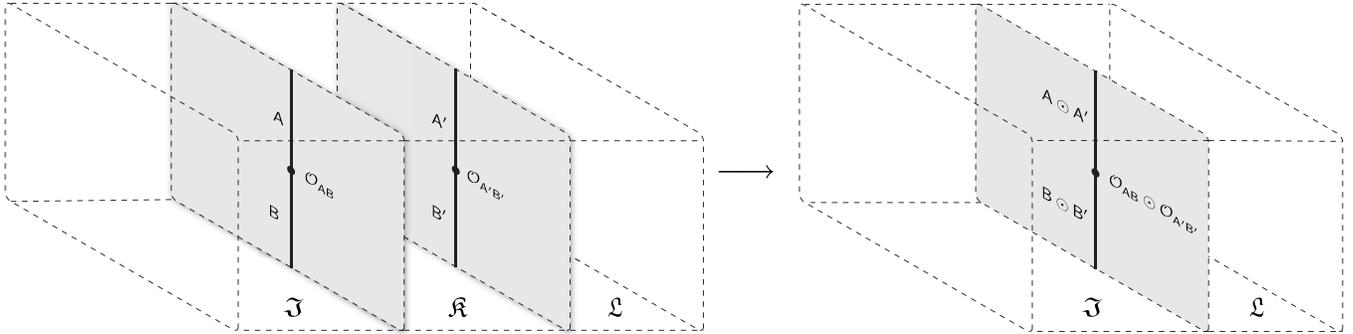}
\caption{Regarding defect lines as 2-morphisms and local operators as 3-morphisms of the 3-category of 3d TFTs gives rise to yet another composition operation between them, which we denote $\odot$. }
\end{figure}
Altogether, one can summarize the situation by saying that 3d TFTs form a 3-category. Its objects are 3d TFTs, its morphisms are walls between 3d TFTs, its 2-morphisms (morphisms between morphisms) are defect lines on walls, and its 3-morphisms (morphisms between 2-morphisms) are local operators sitting on defect lines. 

If we restrict attention to surface operators in a particular 3d TFT, we get a 3-category with a single object. Equivalently, the 2-category of surface operators in a 3d TFT has an extra structure which allows one to fuse objects, morphisms and 2-morphisms. In other words, it is a monoidal 2-category. It has an identity object (the trivial surface operator) whose endomorphisms can be thought of as defect lines in the bulk. 

To determine the category of bulk defect lines in a given 3d TFT one may use the dimensional reduction trick. To apply it, we excise a tubular neigborhood of a defect line and replace the defect line by a suitable boundary condition on its boundary. The collar neighborhood of the newly created boundary locally looks like $S^1\times\RR_+\times \RR$, where $\RR_+$ corresponds to the ``radial'' direction. Therefore we may identify the defect line with the boundary condition in the 2d TFT obtained by compactifying the 3d TFT on a circle. This trick allows one to determine the category of bulk defect lines by studying the category of branes in a 2d TFT. We lose some information in this way: the category of bulk defect lines in a 3d TFT is in fact a braided monoidal category (i.e. a category with a quasi-commutative tensor product), but this structure cannot be seen from the 2d viewpoint. 

Similarly, the study of categories of boundary defect lines reduces to the study of the category of branes in a 2d TFT obtained by compactifying the 3d TFT on an interval. If the boundary defect line separates two boundary conditions $\X$ and $\Y$, then the boundary conditions on the endpoints of the interval should be $\X$ and $\Y$. This 2d viewpoint entails some information loss: for example, it does not allow one to compute the monoidal structure on the category $\dV_{\X\X}$.

\subsection{Four-dimensional TFT and 2-categories of surface operators}

Boundary conditions in a 4d TFT form a 3-category. For example, if the 4d TFT is trivial, its 3-category of boundary conditions is the 3-category of all 3d TFTs. Similarly, walls (codimension-1 defects) in a given 4d TFT form a monoidal 3-category. In this paper we will avoid dealing with such complicated structures and will focus instead on defects of codimension two, i.e. surface operators in a 4d TFT. Such surface operators form a 2-category. One way to see it is to apply the dimensional reduction trick: excise a tubular neighborhood of a surface operator and replace the operator by a suitable boundary condition on the newly created boundary. The collar neighborhood of the boundary looks locally like $S^1\times \RR_+\times\RR^2$, so we may reinterpret a surface operator as a boundary condition in a 3d TFT obtained by compactifying the 4d theory on a circle. Then we can appeal to the known fact that boundary conditions in a 3d TFT form a 2-category.

Of course, one can also explain the meaning of this 2-category structure directly in 4d terms. The type of a surface operator may jump across a defect line, and one can regard defect lines on surface operators as morphisms in a 2-category. Local operators sitting at a junction point of two surface defect lines are 2-morphisms. The 4d viewpoint also makes it clear that the 2-category of surface operators has a rich extra structure. First of all, one may fuse surface operators together with defect lines and local operators sitting on them. This gives rise to a monoidal structure on the 2-category of surface operators. Second, by moving surface operators around one can easily see that the fusion operation is quasi-commutative, i.e. one gets a braided monoidal 2-category.\footnote{A possible mathematical definition of a braided monoidal 2-category is spelled out in \cite{KapVoe}. However, it appears that braided monoidal structures which arise in 4d TFT are of a rather special kind. In particular, the braiding is always invertible.}

In this paper we will use the dimensional reduction trick to study the 2-category of surface operators and leave the understanding of the braided monoidal structure on this 2-cateory for future work.

\section{Review of the GL-twisted theory}

The bosonic fields in the GL-twisted 4d theory are a gauge field $A_\mu$ (a connection on a principal $G$-bundle $\P$ over a 4-manifold $M_4$), a 1-form $\phi_\mu dx^\mu$ with values in $\Ad(\P)$, and a 0-form $\sigma$ with values in the complexification of $\Ad(\P)$. The conventions are the same as in \cite{KW}; in particular, real adjoint-valued fields are regarded as anti-Hermitian, and the covariant derivative in the adjoint representation takes the form $d_A=d+[A,\cdot]$. The fermionic fields are a pair of $\Ad(\P)_\CC$-valued 1-forms $\psi$ and $\tpsi$, a pair of $\Ad(\P)_\CC$-valued 0-forms $\eta$ and $\teta$, and an $\Ad(\P)_\CC$-valued 2-form $\chi$. The fields $A$ and $\phi$ have ghost number $0$, the fields $\psi$ and $\tpsi$ have ghost number $1$, the fields $\eta,\teta,$ and $\chi$ have ghost number $-1$, and the field $\sigma$ has ghost number $2$. The BRST transformations are
\begin{align*}
\delta A &=i(\psi+t\tpsi),\\
\delta\phi & =i(t\psi-\tpsi),\\
\delta\sigma &=0,\\
\delta\bsigma &=i(\eta+t \teta),\\
\delta \psi & = d_A\sigma+t[\phi,\sigma]\\
\delta\tpsi & = t d_A\sigma -[\phi,\sigma],\\
\delta\eta & = t d_A^*\phi+[\bsigma,\sigma],\\
\delta\teta & = - d_A^*\phi+t[\bsigma,\sigma],\\
\delta\chi & =\frac{1+t}{2}(F-\frac12 [\phi,\phi]+ * d_A\phi)+\frac{1-t}{2}(* (F-\frac12 [\phi,\phi])-d_A\phi).
\end{align*}
Here $t$ takes values in $\CC\bigcup \{\infty\}$, $\bar\sigma=-\sigma^\dagger$, $*$ is the 4d Hodge star operator, and $d^*=* d *$. For $t\neq \pm i$ the action can be wriiten as a BRST-exact term plus a term which depends only on the topology of the bundle $\P$:
\begin{equation}
S=\delta \int_{M_4} V-\frac{\Psi}{4\pi i}\int_{M_4} \Tr\, F\wedge F,
\end{equation}
where
$$
\Psi=\frac{\theta}{2\pi}+\frac{4\pi i}{e^2}\frac{t^2-1}{t^2+1}.
$$
Here $\theta$ is the theta-angle of the 4d gauge theory and $e^2$ is the gauge coupling.
The explicit form of $V$ can be found in \cite{KW}.

The simplest surface operators have been introduced by Gukov and Witten \cite{GW}. They are disorder operators corresponding to a codimension-2 singularity in the fields of the form
$$
A=\alpha d\theta, \quad \phi=\beta \frac{dr}{r}-\gamma d\theta.
$$
Here $\alpha$ is an element of a maximal torus $\mT$ of $G$, and $\beta,\gamma$ are elements of the Lie algebra $\frt$ of $\mT$.  For simplicity, let us assume that the triple $(\alpha,\beta,\gamma)$ breaks $G$ down to $\mT$. Gauge transformations which preserve $\mT$ form the Weyl group $\cW$; the triplet $(\alpha,\beta,\gamma)$ is defined up to the action of $\cW$ on $\mT \times \frt\times\frt$. All fields other than $A$ and $\phi$ are nonsingular. 

The surface operator depends on an additional parameter $\eta$ taking values in the torus $\Hom(\Lambda_\coch,U(1))$. Here $\Lambda_\coch$ is the lattice of magnetic charges $\Hom(U(1),\mT)$. Equivalently, as explained in \cite{GW}, $\eta$ can be thought of as taking values in $\mTL$, the maximal torus of the Langlands-dual group. The parameter $\eta$ arises as follows. First, note that the above singularity in the fields breaks the gauge group down to $\mT$. Thus if $D$ is the codimension-2 submanifold on which the surface operator is supported, the restriction of the gauge field to $D$ has a first  Chern class $c_1\vert_D$  taking values in $\Lambda_\coch$. Given $\eta$ we can insert into the path-integral a phase factor
$$
\eta(c_1(D)).
$$
This factor depends only on the behavior of the gauge field on $D$ and can be regarded as an $\eta$-dependent  modification of the surface operator defined above. 

Gukov-Witten surface operators are BRST-invariant for arbitrary $t$, but their properties depend on $t$. We will see below that there are many other surface operators. In what follows we will focus on the cases $t=i$, $t=1$, and $t=0$. The first two cases are exchanged by S-duality (at zero $\theta$-angle) and play a prominent role in the physical approach to the Geometric Langlands Program \cite{KW}. The last case is self-dual and is the most natural starting point for understanding Quantum Geometric Langlands Duality \cite{Kap:qGL}. 

\section{Surface operators at $t=i$: the abelian case}

\subsection{Reduction to 3d}

As explained in \cite{GW}, at $t=i$ varying the parameters $\beta$ and $\eta$ changes the surface operator only by BRST-exact terms. Thus Gukov-Witten operators depend on a single complex parameter $\alpha-i\gamma$. But there exist much more general surface operators. To study them systematically, it is convenient to use the fact that surface operators in the 4d TFT are in 1-1 correspondence with boundary conditions in the 3d TFT compactified on a circle. The advantage of the 3d viewpoint is that the problem of classification of boundary conditions is more familiar. In particular, for $t=i$ the 3d TFT that one gets is a gauged version of the Rozansky-Witten model, so we can use many of the results of \cite{KRS} where boundary conditions for the Rozansky-Witten model have been studied.

In this section we consider the case $G=U(1)$. Reduction to 3d amounts to declaring all fields to be independent of the $x^4$ direction which is periodic with period $2\pi$. 
The reduced theory has the following bosonic fields: a 3d gauge field $A$, a 1-form $\phi$, a complex  0-form $\sigma$, and a pair of 0-forms $A_4$ and $\phi_4$. More properly, one should work with a $U(1)$-valued scalar $\exp(-2\pi A_4)$ which represents the holonomy of the gauge field along the compact direction. This field is invariant with respect to $x^4$-dependent gauge transformations
$$
A_4\mapsto A_4 + im,\quad m\in\ZZ.
$$
The fermionic fields are 1-forms $\psi,\tpsi,\chi,\tchi$, and 0-forms $\eta,\teta,\psi_4,\tpsi_4$.

At $t=i$ it is convenient to combine $A_4$ and $\phi_4$ into a complex 0-form $\tau=A_4+i\phi_4$, or more properly into a gauge-invariant $\CC^*$-valued scalar 
$\exp(-2\pi\tau)$. Then $\tau$ and $\sigma$ are BRST-invariant. We also define the complex 3d gauge field $\cA=A+i\phi$ which is BRST-invariant and the corresponding curvature $\cF=d\cA$. The BRST transformations of other fields are
\begin{align*}
\delta (A-i\phi) &=2i(\psi+i\tpsi),\\
\delta\bsigma &=i(\eta+i \teta),\\
\delta \tau & =-2i(\psi_4+i\tpsi_4) ,\\
\delta \psi & = d\sigma\\
\delta\tpsi & = i d\sigma,\\
\delta\psi_4 &= 0,\\
\delta\tpsi_4 &=0,\\
\delta\eta & = i d^\star\phi,\\
\delta\teta & = - d^\star\phi,\\
\delta\chi & = \cF,\\
\delta\tchi& =d\tau.
\end{align*}
Here $d^\star=\star d\star$ and $\star$ denotes the 3d Hodge star operator.

These BRST-transformations are nilpotent off-shell. One can make them nilpotent on-shell by introducing a suitable auxiliary field, as discussed in the appendix E.  The 3d action contains both a BRST-exact metric-dependent term and a BRST-closed metric-independent term. Its explicit form is given in the appendix E. 

The analysis of boundary conditions in the 3d theory is greatly facilitated by the observation that this 3d theory decomposes into two independent sectors, the Rozansky-Witten model with target $T^*\CC^*$ and a topological $U(1)$ gauge theory. Let us discuss these two 3d TFTs in turn. 

\subsection{Rozansky-Witten model with target $T^*\CC^*$}

The fields of this model are a subset of the fields of the 3d theory listed above. The bosonic ones are the $\CC^*$-valued scalar $h=\exp(-2\pi \tau)$ and the $\CC$-valued scalar $\sigma$. The fermionic ones are the 0-forms $\psi_4+i\tpsi_4,\eta+i\teta$ and the 1-forms $\psi-i\tpsi,\tchi$. The RW model can be defined for any complex symplectic target space $X$, and $T^*\CC^*$ is a special case with the symplectic form $d\tau\wedge d\sigma$. It is shown in appendix E that the correct 3d action arises from the 4d action of the GL-twisted theory upon reduction.

For a general $X$ the RW model has $\ZZ_2$ ghost number symmetry, but as explained in \cite{KRS} when $X$ is a cotangent bundle one can promote it to a $U(1)$ ghost number symmetry by letting the fiber coordinates have ghost number two. This agrees with the fact that $\sigma$ has ghost number two already in the 4d theory. To emphasize that the fiber coordinate has ghost number two we will denote the target manifold $T^*[2]\CC^*$.

According to \cite{KRS} the simplest boundary conditions in the RW model correspond to complex Lagrangian submanifolds of $X$. If we want to preserve ghost number symmetry, these Lagrangian submanifolds must be invariant with respect to the rescaling $\sigma\mapsto \lambda^2\sigma$, $\lambda\in\CC^*$. This requires the Lagrangian submanifold of $T^*\CC^*$ to be the conormal bundle of a complex submanifold in $\CC^*$. This means that a (closed) $\CC^*$-invariant complex Lagrangian submanifold is either the zero section $\sigma=0$ or one of the fibers of the cotangent bundle given by $\tau=\tau_0$. The zero section boundary condition plays a special role and will be denoted $\X_0$ in this subsection.

More general boundary conditions correspond to families of B-models or Landau-Ginzburg models parameterized by points in a complex Lagrangian submanifold. As mentioned in \cite{KRS} and explained in more detail in \cite{KR} it is sufficient to restrict oneself to the case when the Lagrangian submanifold is the zero section $\sigma=0$. One can describe these boundary conditions more algebraically as follows. Recall that the category of boundary line operators on the boundary $\X_0$ is a monoidal category which we denote $\dV_{\X_0\X_0}$. Given any boundary condition $\X$ one may consider the category $\dV_{\X\X_0}$ of boundary defect lines which may separate $\X$ from $\X_0$. This category is a module category over the monoidal  category $\dV_{\X_0\X_0}$. It was proposed in \cite{KRS} that this module category completely characterizes the boundary condition $\X$. Concretely, in the case of the RW model with target $T^*[2]\CC^*$ the category of boundary line operators $\dV_{\X_0\X_0}$ is equivalent to $D^b(Coh(\CC^*))$. One way to see it is to reduce the RW model on an interval with the boundary condition $\X_0$ on both boundaries. The resulting 2d TFT is a B-model with target $\CC^*$, and its category of branes may be identified with $D^b(Coh(\CC^*))$. The 2d viewpoint does not allow one to determine the monoidal structure, but one can show that it is given by the usual derived tensor product \cite{KRS,KR}.

It was further argued in \cite{KRS,KR} that the 2-category of boundary conditions for the RW model with target $T^*[2]\CC^*$ is equivalent to the derived 2-category of module categories over $D^b(Coh(\CC^*))$. That is, it is the 2-category of derived categorical sheaves over $\CC^*$ as defined by B.~Toen and G.~Vezzosi \cite{ToVe}. This provides an  algebraic description of boundary line operators and their OPEs for all boundary conditions.

\subsection{B-type topological gauge theory with gauge group $U(1)$ }

There are two different topological gauge theories in 3d which can be obtained by twisting $N=4$ $d=3$ super-Yang-Mills theory. The first one is the dimensional reduction of the Donaldson-Witten twist of $N=2$ $d=4$ super-Yang-Mills theory. The second one is intrinsic to 3d and has been first discussed by Blau and Thompson \cite{BT}. We will refer to them as A-type and B-type topological gauge theories respectively. The reason for this terminology is that the BPS equations in the former theory are elliptic, as in the usual A-model, while in the latter theory they are overdetermined, as in the usual B-model. The definition and some properties of the B-type 3d gauge theory (for a general gauge group) are described in the appendix C. In this subsection we only deal with the abelian case.

Consider the 3d bosonic fields $A$, $\phi$ and the fermionic fields $\psi+i\tpsi,\chi,\eta-i\teta,\psi_4-i\tpsi_4$. It is easy to check that their BRST transformations at $t=i$ are exactly the same as for the B-type 3d gauge theory. The action has a BRST-exact metric-dependent piece and a BRST-closed metric-independent piece:
$$
S=-\frac{1}{2e^2}\, \delta\int_{M_3} \left(\chi\wedge \star \cF-\frac{i}{2}(\eta-i\teta)\wedge \star d^\star\phi\right)+\frac{1}{2 e^2}\int_{M_3}(\psi_4-i\tpsi_4) d\chi
$$
In principle we should gauge-fix the theory and modify the BRST operator appropriately; we leave this as an exercise for the reader. 

As in any gauge theory, the most natural boundary conditions are the Dirichlet and Neumann ones. The Dirichlet condition requires the restriction of $A+i\phi$ to the boundary to be trivial. In addition, one requires $\phi_3$ (the component of $\phi$ orthogonal to the boundary) to satisfy the Neumann condition $\partial_3\phi_3=0$. BRST-invariance then fixes the boundary conditions for fermions: the restriction of the forms $\psi+i\tpsi$, $\chi$ and $\eta-i\teta$ to the boundary must vanish, The Neumann boundary condition leaves the restriction of $\cA$ to the boundary unconstrained but requires the restriction of the 1-form $\star\cF=\star d\cA$ to vanish. In addition $\phi_3$ must satisfy the Dirichlet boundary condition, i.e. it must take a prescribed value on the boundary. In the Neumann case BRST-invariance requires the restrictions of  the fermions $\star\chi$, $\psi_3+i\tpsi_3$ and $\psi_4-i\tpsi_4$ to vanish. Note that in the Dirichlet case the gauge group is completely broken at the boundary, while in the Neumann case it is unbroken. 

The Dirichlet condition does not have any parameters, while the Neumann condition seems to depend on a single real parameter $\beta$, the boundary value of $\phi_3$. 
On the quantum level there is another parameter: we can add to the action a boundary topological term 
$$
\theta \int_{\partial M_3} \frac{\cF}{2\pi}
$$
In fact, both parameters are irrelevant, in the sense that topological correlators do not depend on them. The irrelevance of the parameter $\theta$ follows from the fact that the above topological term is BRST-exact and equal to
$$
\frac{\theta}{2\pi}\delta  \int_{\partial M_3} \chi
$$
To see the irrelevance of the parameter $\beta$, note that to shift $\beta$ we need to add to the action a boundary term proportional to
$$
\int_{\partial M_3} \partial_3\phi_3
$$
Since $\phi_1$ and $\phi_2$ vanish on the boundary, this is also BRST-exact and proportional to
$$
\delta \int_{\partial M_3} (\eta-i\teta).
$$

Following the same line of thought as in \cite{KRS}, one can try to describe the 2-category of boundary conditions in this theory by picking a distinguished boundary condition $\X_0$ and  characterizing any other boundary condition $\X$ by the category $\dV_{\X\X_0}$ of defect line operators between $\X$ and $\X_0$. That is, one attaches to any boundary condition $\X$ a module category $\dV_{\X\X_0}$ over the monoidal category $\dV_{\X_0\X_0}$.

An obvious guess for the distinguished boundary condition is the free (Neumann) one since it leaves the gauge group unbroken. To determine the category of boundary line operators $\dV_{\X_0\X_0}$ for this boundary condition, one may reduce the 3d theory on an interval and study the category of branes in the resulting 2d TFT. In the Neumann case, reduction on an interval gives the following result: the bosonic fields are the gauge field $A$ and the 1-form $\phi$, the fermionic ones are the 0-form $\eta-i\teta$, the 1-form $\psi+i\tpsi$ and the 2-form $\chi$. This is the field content of a B-type topological gauge theory in 2d, see appendix A. It is easy to check that the BRST transformations of these fields are also the same as in the B-type 2d gauge theory. The category of branes for this 2d TFT is the category of graded finite-dimensional representations of $G=U(1)$, see appendix A for details. This is because the only boundary degrees of freedom one can attach are described by a vector space which carries a representation of the gauge group. The monoidal structure cannot be determined from 2d considerations, but it easy to see that it is given by the usual tensor product. Indeed, as described in the appendix, a brane corresponding to a representation space $V$ is obtained by inserting the holonomy of the complex connection $\cA=A+i\phi$ in the representation $V$ into the path-integral. From the 3d viewpoint this means that the corresponding boundary line operator is the Wilson line operator for $\cA$ in the representation $V$. On the classical level, the fusion of two Wilson line operators in representations $V_1$ and $V_2$ gives the Wilson line in representation $V_1\otimes V_2$, and clearly there can be no quantum corrections to this result (the gauge coupling $e^2$ is an irrelevant parameter). 

To summarize, the monoidal category $\dV_{\X_0\X_0}$ is the category of graded finite-dimensional representations of $\CC^*$, or equivalently the equivariant derived category of coherent sheaves over a point which we denote $D^b_{\CC^*}(Coh(\bul))$. We propose that the 2-category of boundary conditions is equivalent to the 2-category of module categories over $D^b_{\CC^*}(Coh(\bul))$. To give a concrete class of examples of such a module category, consider a Calabi-Yau manifold $Y$ with a $\CC^*$ action. The corresponding B-model can be coupled to the boundary gauge field and provides a natural set of topological boundary degrees of freedom for the 3d gauge theory. The corresponding category of boundary-changing line operators is the $\CC^*$-equivariant bounded derived category of $Y$ which is obviously a module category over $D^b_{\CC^*}(Coh(\bul))$.

\subsection{Putting the sectors together}

It is fairly obvious how to combine the two models. The most basic boundary condition in the full theory is  $\sigma=0$ in the RW sector and the free (Neumann) condition in the gauge sector.  We will call this the distinguished boundary condition.  The bosonic fields which are free on the boundary are the $\CC^*$-valued scalar $h=\exp(-2\pi\tau)$ and the restriction of the complex gauge field $\cA=A+i\phi$. More general boundary conditions involve a boundary B-model or a boundary Landau-Ginzburg model fibered over $\CC^*$ and admitting a $\CC^*$-action. The fibration over $\CC^*$ determines the coupling to the boundary value of $\tau$, while the $\CC^*$-action determines the coupling to the boundary gauge field $\cA$. 

As in the RW model, we can give a more algebraic definition of the set of all boundary conditions in the full theory. This description is also useful because it suggests how to define the 2-category structure of the set of boundary condtions. We consider the monoidal category of boundary line operators for the distinguished boundary condition. 
This is the category of branes for the 2d TFT obtained by reducing the gauged RW model on an interval. Since the reduction of the B-type 3d gauge theory gives the B-type 2d gauge theory, and the reduction of the RW model gives the B-model with target $\CC^*$, the effective 2d TFT is the gauged B-model with target $\CC^*$, where the gauge group $U(1)$ acts trivially. As described in appendix B, the corresponding category of branes is equivalent to $D^b_{\CC^*}(Coh(\CC^*))$. The monoidal structure cannot be determined from the 2d considerations, but it is easy to see (given the results for the RW model and the B-type gauge theory in 3d) that it is given by the derived tensor product.

Every boundary condition gives rise to a module category over this monoidal category. It is natural to conjecture that the converse is also true, i.e. every reasonable module category over this monoidal category can be thought of as a boundary condition for the full 3d TFT. For example, we may consider a family of Calabi-Yau manifolds parameterized by points of $\CC^*$ such that each model in the family has a $\CC^*$ symmetry. The corresponding module category is the $\CC^*$-equivariant derived category of the total space of the fibration. This gives us a conjectural description of the 2-category of surface operators in the parent 4d gauge theory.

Let us describe how Gukov-Witten surface operators fit into this picture. Such operators depend on a complex parameter $h_0=\exp(-2\pi (\alpha-i\gamma))$ taking values in $\CC^*$. 
From the 3d viewpoint, $h_0$ determines the boundary value of the scalar $h=\exp(-2\pi\tau)$ in the RW sector. The other scalar $\sigma$ is left free. Thus the boundary conditions for the RW sector correspond to a Lagrangian submanifold of $T^*[2]\CC^*$ given by $h=h_0$ (the fiber over the point $h_0$). The gauge sector boundary conditions are of Neumann type and have no nontrivial parameters.There are no boundary degrees of freedom. From our algebraic viewpoint we may describe this as follows. In the usual RW theory the fiber over $h=h_0$ corresponds to a skyscraper sheaf of DG-categories over $\CC^*$ whose ``stalk'' over $h_0$ is the category of bounded complexes of vector spaces. We may denote it $D^b(Coh(\bul))$. Including the gauge degrees of freedom means working with a sheaf of categories with a $\CC^*$ action. Thus we simply consider a skyscraper sheaf of categories over $\CC^*$ whose ``stalk'' over $h_0$ is the category of  $\CC^*$-equivariant complexes of vector spaces $D^b_{\CC^*}(Coh(\bul))$. The monoidal category $D^b_{\CC^*}(Coh(\CC^*))$ acts on it in a fairly obvious manner: one simply tensors an object of $D^b_{\CC^*}(Coh(\bul))$ with the (derived) restriction of an object of $D^b_{\CC^*}(Coh(\CC^*))$ to the point $h=h_0$.

\subsection{Line operators on Gukov-Witten surface operators}

The category of morphisms between two different skycraper sheaves of categories is trivial (the set of objects is empty). This corresponds to the fact that two different Gukov-Witten surface operators cannot join along a boundary-changing line operator. But the category of line operators sitting on a particular Gukov-Witten surface operator (i.e. the endomorphism category of a Gukov-Witten surface operator) is nontrivial. Its most obvious objects are Wilson lines for the complexified gauge field $\cA$, which are obviously BRST-invariant. Such operators are labeled by irreducible representations of $\CC^*$. One might guess therefore that the category of surface line operators is simply the category of representations of $\CC^*$, or perhaps the category of $\CC^*$-equivariant complexes of vector spaces which we denoted $D^b_{\CC^*}(Coh(\bul))$ above. However, this naive guess is wrong, which can be seen by inspecting BRST-invariant local operators which can be inserted into such a Wilson line operator. From the abstract viewpoint they form an algebra (the endomorphism algebra of an object in the category of line operators). It is clear that any power of the field $\sigma$ gives such an operator, so the algebra of local operators on a line operator is the algebra of polynomial functions of a single variable of ghost number $2$. In what follows we will denote the line parameterized by $\sigma$ by $\CC[2]$ to indicate that $\sigma$ sits in degree $2$; thus $\CC[2]$  is a purely even graded manifold. On the other hand, the algebra of endomorphisms of an irreducible representation of $\CC^*$ is simply $\CC$. 

To determine what the category of line operators is it is convenient to take the 2d viewpoint and reduce the 3d theory on an interval with the Gukov-Witten-type boundary condition on both ends. Let $x^3$ denote the coordinate on the interval. Gukov-Witten boundary conditions eliminate the complex scalar $h$ (which is now locked at the value $h_0$) and the field $\phi_3$ but keep the complex scalar $\sigma$ and the gauge field $\cA$. Thus the effective 2d theory also has two sectors: the B-model with target $\CC[2]$ and the B-type 2d topological gauge theory. According to appendix B, the corresponding category of branes is equivalent to the $\CC^*$-equivariant derived category of $\CC[2]$: its objects can be regarded as $\CC^*$-equivariant complexes of holomorphic vector bundles on $\CC[2]$ (with a trivial $\CC^*$ action on $\CC[2]$). 

This answer is independent of the parameter $h_0=\exp(-2\pi(\alpha-i\gamma))$ of the Gukov-Witten surface operator.  In particular, we can choose the trivial surface operator $h_0=1$, in which case we should get the category of bulk line operators in the GL-twisted theory at $t=i$. 

It is not difficult to see that this answer for the category of bulk line operators agrees with the computation of the endomorphism algebra of a Wilson line explained above. Indeed, an insertion of a Wilson line does not put any constraints on $\sigma$ and does not add any degrees of freedom, and therefore should correspond to a trivial line bundle over $\CC[2]$. Its fiber carries a representation of $\CC^*$ determined by the charge of the Wilson line. The endomorphism algebra of such an object of $D^b_{\CC^*}(Coh(\CC[2]))$ is simply the algebra of polynomial functions on $\CC[2]$. 

It is now clear that the category of line operators contains objects other than Wilson lines. For example, we may consider a skyscraper sheaf at the origin of $\CC[2]$, whose stalk at the origin is a complex line $V$ carrying some representation of $\CC^*$. There are two different way to define the corresponding line operator. First, we may consider a free resolution of the skyscraper:
$$
V[-2] \otimes \cO \ra  V\otimes \cO,
$$
where $V[-2]$ means $V$ placed in ghost degree $-2$, $\cO$ is the algebra of polynomial functions on $\CC[2]$, and the cochain map is multiplication by $\sigma$.  The shift by $-2$ is needed so that the cochain map has total degree $1$. The existence of such a resolution means that we can realize the ``skyscraper'' line operator as a ``bound state'' of two Wilson lines both associated with the representation $V$ but placed in different cohomological degrees. The corresponding bulk line operator is obtained using the formulas of appendix 2, where the target of the gauged B-model is taken to be $\CC[2]$, the vector bundle $E$ on $\CC[2]$ is trivial and of rank $2$, with graded components in degrees $1$ and $0$, and the bundle morphism $T$ from the former to the latter component is multiplication by $\sigma$. In accordance with the appendix, we consider a superconnection on $\sigma^*E$ of the form
$$
\cN=\begin{pmatrix} n\cA & 0\\  \frac12 (\psi-i\tpsi) &  n\cA\end{pmatrix}, 
$$
where $n\in\ZZ$ is the weight with which $\CC^*$ acts on $V$. The bulk line operator corresponding to the skyscraper sheaf at the origin of $\CC[2]$ is the holonomy of this superconnection along the insertion line $\ell$.

Another (equivalent) way is to take seriously the fact that the skyscraper sheaf is localized at $\sigma=0$ and require the field $\sigma$ to vanish at the insertion line $\ell$. To make this well-defined, one needs to excise a small tubular neighborhood of $\ell$ and impose a suitable boundary condition on the resulting boundary. This condition must set $\sigma=0$ and leave the components of $\cA$ tangent to the boundary $\ell$ unconstrained.  BRST-invariance determines uniquely the boundary conditions for all other fields.

\section{Surface operators at $t=i$: the nonabelian case}

\subsection{Reduction to 3d in the nonabelian case}\label{sec:nonabelianred}

To generalize the preceding discussion to the nonabelian case we need to understand the 3d TFT which is obtained by compactifying the 4d gauge theory on a circle. This is less straightforward than in the abelian case, because requiring the fields to be independent of the $x^4$ coordinate is not a gauge-invariant condition. One can try to avoid dealing with this issue by first fixing a gauge such that $A_4$ does not depend on $x^4$. This works in the neighborhood of $A_4=0$, i.e. when the holonomy of $A$ along $S^1$ is close to $1$. But in general the condition that $A_4$ is $x^4$-independent does not fix the freedom to make $x^4$-dependent gauge transformations. For example, suppose $A_4$ is proportional to an element $\mu\in\frg$ which satisfies 
$$
\exp(2\pi \mu)=1.
$$
Such $\mu$ are precisely those which lie in the $G$-orbits of the cocharacter lattice of $G$.
Then the gauge transformation
$$
g(x^4)=\exp(\mu x^4)
$$
shifts $A_4$ by $\mu$:
$$
A_4\mapsto A_4+\mu
$$
Such a gauge transformation in general makes other fields $x^4$-dependent. 

It is shown in appendix E that the naive reduction procedure which requires all fields to be independent of $x^4$ gives the gauged Rozansky-Witten model with target $T^*[2]\frg_\CC\simeq \frg_\CC \times \frg_\CC^*[2]$, where the gauge group $G$ acts on the base $\frg$ and the fiber $\frg^*[2]$ via the adjoint and coadjoint representations, respectively. The symplectic form is the canonical form on the cotangent bundle. The true target space of the reduced model is $T^*[2]G_\CC$ which contains an open neighborhood of the origin in $T^*[2]\frg_\CC$ as an open subset. We conjecture that the 3d theory is the gauged Rozansky-Witten model with target $T^*[2]G_\CC$, basically because it is the only obvious possibility.

\subsection{Some simple boundary conditions}

Let us consider some boundary conditions in the gauged Rozansky-Witten model with target $T^*[2]G_\CC$. The most natural boundary condition in the gauge sector is the Neumann condition, which preserves full gauge-invariance on the boundary. In the matter sector one has to pick a $G$-invariant complex Lagrangian submanifold of $T^*[2]G_\CC$ which is invariant with respect to the rescaling of the fiber. Such a Lagrangian submanifold can be constructed by picking a $G_\CC$-invariant closed complex submanifold of $G_\CC$ and taking its conormal bundle. For example, one can take the whole $G_\CC$, and then the Lagrangian submanifold is given by $\sigma=0$. We will call the resulting boundary condition in the gauged RW model the distinguished boundary condition. It is an analogue of the NN condition in the abelian case. 

Another natural choice of a $G$-invariant Lagrangian submanifold is the conormal bundle of a complex conjugacy class in $G_\CC$. In order for the submanifold to be closed take the conjugacy class to be semisimple. This boundary condition is a nonabelian analogue of the ND condition. The corresponding surface operator is a semisimple Gukov-Witten-type surface operator. Indeed, fixing a semisimple conjugacy class of $\exp(-2\pi(A_4+i\phi_4))$ is the same as fixing a semisimple conjugacy class of the limiting holonomy of the complex connection $A+i\phi$ in the 4d gauge theory. More generally, if the conjugacy class is not closed, one needs to consider the conormal bundle of its closure. 

It is easy to analyze boundary line operators for these boundary conditions. Reducing the 3d theory on an interval with the distinguished boundary conditions we get a B-type 2d gauge theory coupled to a B-model with target $G_\CC$. The gauge group acts on $G_\CC$ by conjugation. According to appendix B, the corresponding category of branes is equivalent to $D^b_{G_\CC}(Coh(G_\CC))$. The monoidal structure cannot be deduced from the 2d considerations, but the same analysis as in the usual RW model shows that it is given by the derived tensor product. 

In the Gukov-Witten case we need to fix a semisimple complex conjugacy class $\cC$ in $G_\CC$. Let $N^*\cC$ denote the total space of its conormal bundle in $T^*G_\CC$. Concretely, it is the space of pairs $(g,\sigma)$, where $g\in\cC$ and $\sigma\in\frg_\CC$ satisfies $\Tr\, \sigma\, g^{-1}\delta g=0$ for any $\delta g$ tangent to $\cC$ at $g$. The fiber coordinate $\sigma$ has cohomological degree $2$; to indicate this we will denote the corresponding graded complex manifold $N^*[2]\cC$.  Reduction on an interval in the Gukov-Witten case gives a B-type 2d gauge theory coupled to a B-model whose target is $N^*[2]\cC$. Its category of branes is $D^b_{G_\CC}(Coh(N^*[2]\cC))$. The monoidal structure is given by the derived tensor product.

\subsection{Bulk line operators}

It is interesting to consider the special case of a Gukov-Witten surface operator corresponding to the trivial conjugacy class in $G_\CC$ (i.e. the identity).  This is the trivial surface operator, so the category of 3d boundary line operators in this case can be identified with the category of bulk line operators in the 4d TFT. The conormal bundle of the identity element is simply the dual of the complexified Lie algebra $\frg_\CC$; the group $G_\CC$ acts on it by the adjoint representation. Thus the category of 4d bulk line operators is equivalent to $D^b_{G_\CC}(Coh(\frg^*_\CC[2]))$. In other words, it is the $G_\CC$-equivariant derived category of the graded algebra $\oplus_p {\rm Sym}^p \frg$ where the $p^{\rm th}$ component sits in cohomological degree $2p$. 

In view of this result  it is interesting to consider local operators sitting at the junction of two Wilson loops in representations $V_1$ and $V_2$ of $G$. The corresponding objects of the category $D^b_{G_\CC}(Coh(\frg^*_\CC[2]))$ are free modules over $\frA=\oplus_p {\rm Sym}^p \frg[2]$ of the form $V_1\otimes_\CC \frA$ and $V_2\otimes \frA$, with the obvious $G_\CC$ action. The space of morphisms between them is the space of $G_\CC$-invariants in the infinite-dimensional graded representation
$$
V_1^*\otimes V_2\otimes \frA
$$
Indeed, a BRST-invariant and gauge-invariant junction of two Wilson lines should be an operator in representation $V_1^*\otimes V_2$ constructed out of the complex scalar $\sigma$ taking values in $\frg_\CC$. The space of such operators in ghost number $2p$ is $\Hom_G (\Sym^p \frg,V_1^*\otimes V_2)$, where $\Hom_G$ denotes the space of morphisms in the category of representations of $G$. Summing over all $p$ we get the above answer.

\subsection{More general surface operators}

As in the abelian case, the above examples do not exhaust the set of objects in the 2-category of surface operators. For example, in \cite{Witten:wild} more complicated surface operators have been considered which involve higher-order poles for the complex connection $\cA=A+i\phi$. By analogy with the Rozansky-Witten model we propose that the most general surface operator at $t=i$ (or equivalently, the most general boundary condition in the 3d theory) can be defined as a module category of the monoidal category of boundary line operators for the distinguished boundary condition $\X_0$. As explained above, this monoidal category is $D^b_{G_\CC}(Coh(G_\CC))$. Concretely, this means that the most general surface operator can be obtained by fibering a family of 2d TFTs over $G_\CC$, so that the $G_\CC$ action on the base (by conjugation) lifts to a $G_\CC$ action on the whole family. For example, one may consider a complex manifolds $X$ which is a fibration over $G_\CC$, so that fibers are Calabi-Yau manifolds, and one is given a lift of the $G_\CC$ action on the base (by conjugation) to a $G_\CC$ action on the total space. 

Given any surface operator $\X$, one may construct a module category over $D^b_{G_\CC}(Coh(G_\CC))$ by looking at the category of line operators sitting at the junction of $\X$ and the distinguished surface operator $\X_0$. This category is the category of branes in the 2d TFT obtained by compactifying the 3d TFT on an interval, with the boundary conditions on the two ends given by $\X$ and $\X_0$. Equivalently, one may compactify the 4d TFT on a twice-punctured 2-sphere, with surface operators $\X$ and $\X_0$ inserted at the two punctures. 

For example, if we consider a surface operator defined, as in \cite{Witten:wild}, by a prescribed singularity in the complex connection $\cA$, and take into account that the distinguished surface operator is defined by allowing the holonomy of $\cA$ to be free, we see that the space of vacua of the effective 2d TFT is the moduli space of connections on a punctured disc with the prescribed singularity at the origin. Let us denote this moduli space $\cM$. If in the definition of $\cM$ we divide by the group of gauge transformations which reduce to the identity at some chosen point on the boundary of the disk, then $\cM$ is acted upon by $G_\CC$ and is fibered over $G_\CC$ (the holonomy of $\cA$ along the boundary of the disk). It looks plausible that the effective 2d TFT is the B-model with target $\cM$ coupled to a B-type gauge theory with gauge group $G_\CC$. Its category of branes is a module category over $D^b_{G_\CC}(Coh(G_\CC))$.

\section{Surface operators at $t=1$}

\subsection{Reduction to 3d}

For $t=1$ the 4d TFT compactified on a circle also decomposes into two sectors (gauge and matter), but the analysis of boundary conditions is less straightforward because neither sector has been studied previously. For this reason we will restrict ourselves to the abelian case, which is fairly elementary.  

The gauge sector consists of a 3d gauge field $A$, a real bosonic 0-form $\phi_4$, a complex bosonic 0-form $\sigma$, a fermionic 2-form $\chi=\frac12 \chi_{ij} dx^i dx^j$, a fermionic 1-form $\psi+\tpsi,$, and fermionic 0-forms $\psi_4-\tpsi_4,\eta+\teta$. Their BRST transformations read
\begin{align*}
\delta A &=i(\psi+\tpsi),\\
\delta\phi_4 &=i(\psi_4-\tpsi_4),\\
\delta\sigma &=0,\\
\delta\bsigma &= i(\eta+\teta),\\
\delta (\psi+\tpsi) &=2d\sigma,\\
\delta(\psi_4-\tpsi_4) &=0,\\
\delta(\eta+\teta) &=0,\\
\delta\chi &=F+\star d\phi_4
\end{align*}
On-shell they satisfy $\delta^2=2i\delta_g(\sigma)$, where $\delta_g(\sigma)$ is a gauge transformation with the parameter $\sigma$. The action is BRST-exact:
$$
S_{gauge}=-\frac{1}{2e^2}\,\delta \int_{M_3} \left(\chi\wedge\star (F+\star d\phi_4)+\frac12 (\psi+\tpsi)\wedge \star d\bsigma\right).
$$

The matter sector consists of a real periodic scalar $A_4$, a real bosonic 1-form $\phi$, fermionic 1-forms $\psi-\tpsi,\rho=\chi_{i4}dx^i$, and fermionic 0-forms $\psi_4+\tpsi_4,\eta-\teta$. Their BRST transformations read
\begin{align*}
\delta A_4 &=i(\psi_4+\tpsi_4),\\
\delta \phi &=i(\psi-\tpsi),\\
\delta(\psi-\tpsi) &=0,\\
\delta(\psi_4+\tpsi_4)&=0,\\
\delta\rho &=dA_4+\star d\phi,\\
\delta(\eta-\teta)&=2 d^\star\phi.
\end{align*}
On-shell they satisfy $\delta^2=0$. The matter action is also BRST-exact:
$$
S_{matter}=-\frac{1}{2e^2}\, \delta \int_{M_3} \left(\rho\wedge\star (dA_4+\star d\phi)+\frac12 (\eta-\teta)\wedge \star d^\star\phi\right).
$$

\subsection{Boundary conditions in the gauge sector}

The gauge sector is the dimensional reduction of the Donaldson-Witten 4d TFT \cite{Witten:Donaldson} down to 3d. Above we have called this theory an A-type gauge theory. However, this by itself does not teach us very much, since boundary conditions in this theory have not been discussed previously. Without adding boundary degrees of freedom, the only choices are the Dirichlet and Neumann boundary conditions for gauge fields, with BRST-invariance fixing the conditions on all other fields.

\subsubsection{The Dirichlet condition}

Let us begin with the Dirichlet condition which says that the restriction of $A$ to the boundary is trivial. Since $\delta^2=2i\delta_g(\sigma)$, this makes sense only if $\sigma$ also vanishes on the boundary. BRST-invariance then requires $\eta+\teta$ and the restriction of the 1-form  $\psi+\tpsi$ to vanish. The fermionic equations of motion then require the restriction of $\chi$ to vanish, and the BRST-invariance implies that $\phi_4$ must satisfy the Neumann condition $\partial_3\phi_4=0$, where we assumed that the boundary is given by $x^3=0$. 

The Dirichlet boundary condition has the property that it has no nontrivial local BRST-invariant boundary observables. Indeed, the only nonvanishing BRST-invariant 0-form is $\psi_4-\tpsi_4$, but it is BRST-exact. To analyze boundary line operators, we use the dimensional reduction trick and compactify the 3d theory on an interval with the Dirichlet boundary conditions. The only bosonic fields in the effective 2d theory are the constant mode of $\phi_4$ and the holonomy of $A$ along the interval parameterized by $x^3$. That is, the bosonic fields are a real scalar and a periodic real scalar. The effective 2d TFT is therefore a sigma-model with target $\RR\times S^1$. In fact, it can be regarded as an A-model with target $T^*S^1$. The easiest way to see this is to note that the path-integral of the 3d theory localizes on configurations given by solutions of the Bogomolny equations
$$
F+\star d\phi_4=0.
$$
Upon setting all fields to zero except $A_3$ and $\phi_4$ and assuming that they are independent of $x^3$, this equation becomes
$$
dA_3+\stard d\phi_4=0,
$$
where $\stard$ is the 2d the Hodge star operator.  This is an elliptic equation which can be interpreted as the holomorphic instanton equation, provided we declare  $A_3+i\phi_4$ to be a complex coordinate on the target. Since the action of the 4d theory is BRST-exact, so is the action of the 2d model. This agrees with the well-known fact that the action of an A-model is BRST-exact if the symplectic form on the target space is exact. 

The category of line operators on the Dirichlet boundary is therefore the Fukaya-Floer category of $T^*S^1$ whose simplest objects are Lagrangian submanifolds equipped with unitary vector bundles with flat connections. Since this category arises as the endomorphism category of an object in a 2-category, it must have a monoidal structure, which is not visible from the purely 2d viewpoint. In fact, we do not expect the Fukaya-Floer category of a general symplectic manifold to have a natural monoidal structure. We will argue below that the monoidal structure is induced by the mirror symmetry which establishes the equivalence of the Fukaya-Floer category of $T^*S^1$ with $D^b(Coh(\CC^*))$ and the monoidal structure on the latter category. For now we just note that the base $S^1$ has a distinguished point corresponding to the trivial holonomy of $A$ on the interval. The fiber over this point is a Lagrangian submanifold in $T^*S^1$ and is the identity object with respect to the monoidal structure. The distinguished point allows us to identity $S^1$ with the group manifold $U(1)$. 

\subsubsection{The Neumann condition}

Now let us consider the Neumann condition for the 3d  gauge field $A$. This means that the gauge symmetry is unbroken on the boundary and the restriction of the 1-form $\star F$ vanishes. Then the Bogomolny equation requires $\phi_4$ to have the Dirichlet boundary condition $\phi_4=a=const$, and by BRST-invariance $\psi_4-\tpsi_4$ must vanish at $x^3=0$. 
Fermionic equations of motion imply then that $\psi_3+\tpsi_3$ vanishes as well, and since $\delta(\psi_3+\tpsi_3)=2\partial_3\sigma$, the field $\sigma$ satisfies the Neumann condition. Finally,  the restriction of the 1-form $\star \chi$ to the boundary must vanish, in order for the fermionic boundary conditions to be consistent. Indeed, if $x^1$ is regarded as the time direction, then $(\star \chi)_2$ is canonically conjugate to $\psi_3+\tpsi_3$, so if one of them vanishes, so should the other. Similarly, if $x^2$ is regarded as time, then $(\star\chi)_1$ is canonically conjugate to $\psi_3+\tpsi_3$ and therefore must vanish too. 

In the Neumann case the space of BRST-invariant local observables on the boundary is spanned by powers of the field $\sigma$. To determine the category of boundary line operators one has to reduce the 3d gauge theory on an interval with the Neumann boundary conditions. The bosonic fields of the effective 2d theory are the 2d gauge field and the constant mode of the scalar $\sigma$, the fermionic ones are the 0-form $\eta+\teta$, the 1-form  $\psi+\tpsi$, and the 2-form $\chi$. Their BRST transformations are
\begin{align*}
\delta A &=i(\psi+\tpsi),\\
\delta \sigma &=0,\\
\delta\bsigma &=i(\eta+\teta),\\
\delta(\eta+\teta) &=0,\\
\delta (\psi+\tpsi) & =2d\sigma,\\
\delta \chi &= F.
\end{align*}

This 2d TFT can be obtained from the usual $N=(2,2)$ $d=2$ supersymmetric gauge theory by means of a twist which makes use of the $U(1)_V$ R-symmetry. Since this is the same R-symmetry as that used for constructing an A-type sigma-model, we might call this TFT an A-type 2d gauge theory. As far as we know, its boundary conditions have not been analyzed in the literature previously. It is shown in appendix F that its category of branes is equivalent to the bounded derived category of coherent sheaves on the graded line $\CC[2]$. \footnote{This category is equivalent to the $U(1)$-equivariant constructible derived category of sheaves over a point \cite{BL}.} Again, the 3d origin of this category means that it must have monoidal structure. Here it is given by the usual derived tensor product of complexes of coherent sheaves. The trivial line bundle on $\CC[2]$ is the identity object. From the 3d viewpoint, it corresponds to the ``invisible" line operator on the boundary.

As mentioned above, the Neumann condition depends on a real parameter $a$, the boundary value of the scalar $\phi_4$. On the quantum level there is another parameter which takes values in $\RR/2\pi \ZZ$. It enters as the coefficient of a topological term in the boundary action:
$$
\theta\int_{x^3=0} \frac{F}{2\pi}.
$$
Thus overall the Neumann condition in the gauge sector has the parameter space $\RR\times\SS^1\simeq\CC^*$. 

\subsection{Boundary conditions in the matter sector}

We may impose either Dirichlet or Neumann condition on the periodic scalar $A_4$. Let us discuss these two possibilities in turn.

\subsubsection{The Dirichlet condition}

If $A_4$ satisfies the Dirichlet condition, then BRST-invariance requires the 1-form $\phi$ to satisfy the Neumann condition. This means that the components of $\phi$ tangent to the boundary are free and satisfy $\partial_3\phi_1=\partial_3\phi_2=0$, while the component $\phi_3$ takes a fixed value $\phi_3=a$ on the boundary. BRST-invariance also requires the following fermions to vanish on the boundary: $\psi_4+\tpsi_4$, $\psi_3-\tpsi_3$, $\rho_1,$ $\rho_2$. The real parameter $a$ together with the boundary value of $A_4$ combine into a parameter taking values in $S^1\times\RR$. These parameters are actually irrelevant, in the sense that topological correlators do not depend on them. To see this, note that shifting the boundary value of $A_4$ can be achieved by adding a boundary term to the action of the form
$$
\int_{x^3=0} \partial_3 A_4 d^2x =\int_{x^3=0} \left(\delta\rho_3-(\partial_1\phi_2-\partial_2\phi_1)\right) d^2x
$$
We see that up to a total derivative this boundary terms is BRST-exact, hence does not affect the correlators. A similar argument can be made for the boundary value of $\phi_3$.

The reduction on an interval with the Dirichlet boundary conditions gives rise to a 2d TFT whose only bosonic field is a real 1-form $\phi$. Such a 2d TFT has not been considered previously, but it is closely related to an A-model with target $T^*\RR$. To see this, consider an $N=(2,2)$ supersymmetric sigma-model with target $\CC$ (with the standard flat metric). This model has a $U(1)$ symmetry which acts on the target space coordinate $Z$ by
$$
Z\mapsto e^{i\alpha} Z.
$$
One can add a multiple of the corresponding $U(1)$ current to the standard R-current, thereby defining a new R-current. When performing the A-twist, we can choose this modified R-current instead of the standard one. If $Z$ has charge two with respect to the modified R-symmetry, after twist ${\rm Re}\ Z$ and ${\rm Im}\ Z$ will become components of a 1-form. We will call the resulting 2d TFT the modified A-model.

Apart from the bosonic 1-form $\phi$, the modified A-model has a fermionic 1-form $\psi-\tpsi$ and a pair of fermionic 0-forms $\eta-\teta$ and $\rho$ (the latter comes from the component $\rho_3$ of the 1-form $\rho$ in 3d). Their BRST transformations are
\begin{align*}
\delta\phi &=i(\psi-\tpsi),\\
\delta (\psi-\tpsi) &=0,\\
\delta(\eta-\teta) &=2d^\stard\phi,\\
\delta\rho &=\stard d\phi
\end{align*}
Here $\stard$ is the 2d Hodge star operator, and $d^\stard=\stard d\stard$.

To understand the category of boundary line operators in 3d, we need to describe the category of boundary conditions for the modified A-model. This is fairly straightforward. A natural class of boundary conditions is obtained by imposing on the boundary
$$
\left(a \phi+b \stard \phi\right)\vert_{\partial M_2}=0.
$$
The special cases $b=0$ and $a=0$ correspond to the 2d Dirichlet and Neumann conditions. Since the theory obviously has a symmetry rotating $\phi$ into $\stard\phi$, it is sufficient to consider the Neumann condition $\stard\phi\vert=0$. BRST-invariance requires the restriction of $\stard\psi$ and $\rho$ to vanish on such a boundary. It is easy to see that there are no nontrivial BRST-invariant boundary observables (the only BRST-invariant fermion $\psi$ is BRST-exact), so there is no possibility to couple boundary degrees of freedom in a nontrivial way. This implies that the category of boundary conditions is the same as for a trivial 2d TFT, i.e. the category of complexes of finite-dimensional vector spaces. We may denote it $D^b(Coh(\bul))$. 

There is an important subtlety here related to the fact that the scalar $A_4$ is periodic with period $1$. When reducing on an interval, this means that there are ``winding sectors'', where
$$
\int  dx^3 \partial_3 A_4= n,\quad n\in\ZZ.
$$
This winding is constant along a connected component of the boundary and does not affect the 2d theory in any way. We may incorporate it by introducing an additional integer label on each boundary component which serves as a conserved boundary charge. This is mathematically equivalent to saying that the category of boundary conditions is the category of $\CC^*$-equivariant coherent sheaves over a point $D^b_{\CC^*}(Coh(\bul))$. Objects of this category are complexes of finite-dimensional vector spaces with a $\CC^*$-action, such that the differentials in the complex commute with the $\CC^*$ action. Morphisms are required to preserve the $\CC^*$-action, i.e. to have zero $\CC^*$-charge.

\subsubsection{The Neumann condition}

If $A_4$ satisfies the Neumann condition $\partial_3A_4=0$, then BRST-invariance requires $\phi$ to satisfy the Dirichlet condition . That is, the restriction of $\phi$ to the boundary must vanish, and $\phi_3$ must satisfy $\partial_3\phi_3=0$. This boundary condition does not have any parameters. 

The reduction on an interval gives rise to the A-model with the bosonic fields $A_4$ and $\phi_3$. This can be seen for example by looking at the 3d BPS equation $dA_4+\star d\phi=0$ and restricting to field configurations where $\phi_1=\phi_2=0$ and $A_4$ and $\phi_3$ are independent of $x^3$. For such field configuration the BPS equation becomes the holomorphic instanton equation with target $S^1\times\RR\simeq \CC^*$. From the symplectic viewpoint, $\CC^*$ with its standard K\"ahler form is isomorphic to $T^*S^1$. Thus the category of boundary line operators in this case is the Fukaya-Floer  category of $T^*S^1$.  Since this category arises as the category of boundary line operators in the 3d TFT, it must have a monoidal structure. Although the category appears to be the same as in the gauge sector with the Dirichlet boundary condition, we will see that the monoidal structure is completely different and is induced by the equivalence between (a version of) the Fukaya-Floer category of $T^*S^1$ and the constructible derived category of $S^1$ \cite{ZN}. In particular, the identity object (i.e. the invisible boundary line operator) is different and corresponds to the zero section  of $T^*S^1$ with a trivial rank-1 local system. This illustrates the fact that a monoidal structure on branes in a 2d TFT depends on the way this 2d TFT is realized as a compactification of a 3d TFT on an interval.

\subsection{Electric-magnetic duality}

We are now ready to describe how the 4d electric-magnetic duality acts on various boundary conditions described above. Since for both gauge and matter sectors one can have either Dirichlet or Neumann conditions, there are four possibilities to consider. 

From the 3d viewpoint, 4d electric-magnetic duality amounts to dualizing the 3d gauge field $A$ into a periodic scalar, and simultaneously dualizing the periodic scalar $A_4$ into a 3d gauge field. It is easy to see that electric-magnetic duality applied to the A-type gauge theory gives the Rozansky-Witten model with target $T^*[2]\CC^*$, i.e. it maps the A-type gauge sector to the B-type matter sector. Similarly, it maps the A-type matter sector into the B-type gauge theory (with gauge group $U(1)$).  In other words, electric-magnetic duality reduces to particle-vortex duality done twice.  

The dual of the Neumann condition for a periodic scalar is the Dirichlet condition for the gauge field, and vice-versa. We will use this well-known fact repeatedly in what follows.

\subsubsection{The DD condition}

The first possibility is the Dirichlet condition in both gauge and matter sectors at $t=1$. The Dirichlet condition in the A-type gauge sector maps into a boundary condition in the Rozansky-Witten model with target $T^*[2]\CC^*$ which sets $\sigma=0$ on the boundary and leaves the complex scalar $\tau$ free to fluctuate. The Dirichlet condition in the A-type matter sector is mapped to the Neumann condition in the B-type gauge theory. Note that the Dirichlet condition in the A-type matter sector has two real parameters taking values in $S^1$ and $\RR$.  The former one is mapped to a boundary theta-angle, i.e. a boundary term in the action of the form
$$
\theta\int_{x^3=0} \frac{F}{2\pi}=\theta \int_{x^3=0} \frac{\cF}{2\pi}.
$$
The latter parameter is the boundary value of the field $\phi_3$. Both of these parameters are irrelevant, as discussed in section 4. 

As discussed above, the category of boundary line operators in the A-type 3d gauge theory is the Fukaya-Floer category of $T^*U(1)$.  On the other hand, the category of boundary line operators in the Rozansky-Witten model is $D^b(Coh(\CC^*))$, as explained in \cite{KRS}. These categories are equivalent, by the usual 2d mirror symmetry. 

Let us recall how 2d mirror symmetry acts on some objects in this case. The trivial line bundle on $\CC^*$ is mapped to the fiber over a distinguished point of the base $S^1$. This distinguished point allows us to identify $S^1$ with the group manifold $U(1)$.  More generally, we may consider a holomorphic line bundle on $\CC^*$ with a $\bpartial$-connection of the form
$$
\bpartial+i\lambda\frac{d\bz}{\bz},\quad \lambda\in\CC.
$$
We will denote such a line bundle $\cL_\lambda$. Gauge transformations can be used to eliminate the imaginary part of $\lambda$. They also can shift the real part of $\lambda$ by an arbitrary integer. Thus we may regard the parameter $\lambda$ as taking values in $\RR/\ZZ\simeq S^1$. Mirror symmetry maps $\cL_\lambda$ to a Lagrangian submanifold in $T^*U(1)$ which is a fiber over the point $\exp(2\pi i\lambda)\in U(1)$. 

Applying mirror symmetry to the obvious monoidal structure on $D^b(Coh(\CC^*))$ given by the derived tensor product we get a monoidal structure on the Fukaya category of $T^*U(1)$. The trivial holomorphic line bundle on $\CC^*$, which serves as the identity object in $D^b(Coh(\CC^*))$, is mapped  to the Lagrangian fiber over the identity element of $U(1)$. If we consider two Lagrangian fibers over the points $\exp(2\pi i \lambda_1), \exp(2\pi i\lambda_2)\in U(1)$, their mirrors are line bundles $\cL_{\lambda_1}$ and $\cL_{\lambda_2}$. Their tensor product is a line bundle $\cL_{\lambda_1+\lambda_2}$ whose mirror is the Lagrangian fiber over the point $\exp(2\pi i(\lambda_1+\lambda_2))\in U(1)$. Clearly, this rule for tensoring objects of the Fukaya category makes use of the group structure of $U(1)$, i.e. it is a convolution-type tensor product. 

Another natural class of Lagrangian submanifolds to consider are constant sections of $T^*U(1)$, i.e. submanifolds given by the equation $\phi_4=const$. These submanifolds are circles and may carry a nontrivial flat connection. Thus such A-branes are labeled by points of $\RR\times U(1)\simeq \CC^*$. The mirror objects are skyscraper sheaves on $\CC^*$. The derived tensor product of two skyscrapers supported at different points is obviously the zero object. The derived tensor product of a skyscraper with itself can be shown to be isomorphic to the sum of the skyscraper and the skyscraper shifted by $-1$. That is, it is a skyscraper sheaf over the same point whose stalk is a graded vector space $\CC[-1]\oplus \CC$. Applying mirror symmetry, we see that the tensor product of a section of $T^*U(1)$ with itself must be the sum of two copies of the same section, but with the Maslov grading of one of them shifted by $-1$. We do not know how to reproduce this result without appealing to mirror symmetry, i.e. by computing the product of boundary line operators in the A-type gauge theory.

As discussed above, the category of boundary line operators in the A-type matter sector is the category of branes in a somewhat unusual 2d TFT which is a modification of the A-model with target $T^*\RR$. It was argued above that this category is equivalent to $D^b_{\CC^*}(Coh(\bul))$. This agrees with the B-side, where the reduction on an interval gives a B-type 2d gauge theory.

Putting the gauge and matter sectors together, we see that the DD boundary condition on the A-side is mapped to what we called the distinguished boundary condition on the B-side. The category of boundary line operators for such a boundary condition is  the $\CC^*$-equivariant derived category of coherent sheaves $D^b_{\CC^*}(Coh(\CC^*))$ with its obvious monoidal structure. On the A-side we get a graded version of the Fukaya-Floer category of $T^*U(1)$ where a flat vector bundle over a Lagrangian submanifold has an additional integer grading and morphisms are required to have degree zero with respect to it. This grading arises from the winding number of the periodic scalar $A_4$.

We can also interpret the duality in 4d terms. Indeed, it is easy to see that the DD boundary condition on the A-side arises from a 4d Dirichlet boundary condition at $t=1$, while its dual on the B-side arises from the 4d Neumann condition at $t=i$. Thus electric-magnetic duality exchanges Dirichlet and Neumann boundary conditions in 4d, as expected. The surface operators corresponding to such 4d boundary conditions can be interpreted as follows: we excise a tubular neighborhood of the support of the surface operator and impose the 4d boundary condition on the resulting boundary. In a TFT, such a procedure gives a surface operator (i.e. there is no need to take the limit where the thickness of the tubular neighborhood goes to zero). 

\subsubsection{The NN condition}

This condition is the distinguished boundary condition on the A-side, since the gauge group is unbroken on the boundary, and the periodic scalar $A_4$ is free to explore the whole circle. It is mapped by electric-magnetic duality to the Dirichlet boundary condition for the B-type gauge theory and the boundary condition in the RW model with target $\CC^*$ which fixes the $\CC^*$-valued scalar $\tau$ and leaves $\sigma$ free. Note that both the Neumann boundary condition in the A-type gauge theory and the corresponding boundary condition in the RW model have a parameter taking values in $\CC^*\simeq\RR\times U(1)$. 

Let us compare the categories of boundary line operators. The category of boundary line operators in the A-type gauge theory is the bounded derived category of coherent sheaves $D^b(Coh(\CC[2]))$. The category of boundary line operators in the RW model is also $D^b(Coh(\CC[2]))$. The category of boundary line operators in the A-type matter sector is the Fukaya-Floer category of $T^*S^1$. The category of boundary line operators in the B-type gauge sector is $D^b(Coh(\CC^*))$. Their equivalence is a special case of the usual 2d mirror symmetry. 

But there is more: we expect that the categories of boundary line operators are equivalent as monoidal categories. This is easy to see directly for the RW model with target $\CC^*$ and A-type gauge theory with gauge group $U(1)$. Indeed, in both cases typical objects in the category of boundary line operators are complexes of holomorphic vector bundles which can be represented by Wilson line operators on the boundary for some superconnection on the pull-back vector bundle. In the classical approximation, fusing two such boundary line operators corresponds to the tensor product of  complexes, and there can be no quantum corrections to this result.

It is more complicated to compare the monoidal structures for the other pair of dual theories (B-type gauge theory and A-type matter). We will not attempt to do an independent computation on the A-side but instead describe the monoidal structure on the B-side and then explain what it corresponds to on the A-side. 

Note that since $\CC^*$ is a complex Lie group, the category $D^b(Coh(\CC^*))$ has two natural monoidal structures: the derived tensor product, and the convolution-type product. The former one does not make use of the group structure, while the latter one does. The identity object of the former one is the sheaf of holomorphic functions on $\CC^*$, while for the latter structure it is the skyscraper sheaf at the identity point $1\in\CC^*$. It is the latter monoidal structure which describes the fusion of boundary line operators on the B-side. Indeed, the 3d meaning of the coordinate on $\CC^*$ is the holonomy of the connection $A+i\phi$ along a small semi-circle with both ends on the boundary and centered at the boundary line operator (see figure 8).
\begin{figure}[htbp]  \label{fig:SOGL_boundaryline}
\centering
\includegraphics[height=2in]{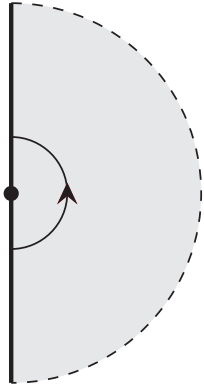}
\caption{A skyscraper sheaf corresponds to a boundary line operator for which the holonomy of $A+i\phi$ along a small semi-circle around it is fixed.  The dot marks the location of the boundary line operator, which we view here in cross-section.}
\end{figure}
Skyscraper sheaves correspond to boundary line operators for which this holonomy is fixed. In particular, the skyscraper sheaf at $1\in\CC^*$ corresponds to the ``invisible'' boundary line operator for which this holonomy is trivial. By definition, this is the identity object in the monoidal category of boundary line operators.

Mirror symmetry maps a skyscraper sheaf on $\CC^*$ to a Lagrangian submanifold of $T^*S^1$ which is a graph of a closed 1-form $\alpha$ on $S^1$. Topologically this submanifold is a circle and is equipped with a trivial line bundle with a flat unitary connection. The moduli space of such an object is $\CC^*$: for $\lambda\in\CC^*$ the phase of $\lambda$ determines the holonomy of the unitary connection, while the absolute value determines the integral of $\alpha$ on $S^1$. Thus the identity object on the B-side is mirror to the zero section of $T^*S^1$ with a trivial flat connection. To describe the monoidal structure on the A-side it is best to recall a theorem of Nadler \cite{Nadler} according to which (a version of) the Fukaya-Floer category of $T^*X$ is equivalent to the constructible derived category of $X$.  Recall that a constructible sheaf on a real manifold $X$ is a sheaf which is locally constant on the strata of a Whitney stratification of $X$; such sheaves can be regarded as generalizations of flat connections. Objects of the constructible derived category are bounded complexes of sheaves whose cohomology sheaves are constructible. The constructible derived category has an obvious monoidal structure arising from the tensor product of complexes of sheaves. The sheaf of locally-constant functions is the identity object with respect to this monoidal structure. According to \cite{ZN,Nadler}, this object corresponds to the zero section of $T^*S^1$ with a trivial flat connection. This suggests that the monoidal structure on the A-side is given by the tensor product on the constructible derived category. It is easy to check that this is compatible with the way mirror symmetry acts on the skyscraper sheaves on $\CC^*$. 

We can try put the gauge and matter sectors together. On the B-side, we have the B-model with target $\CC^*\times \CC[2]$ whose category of branes is $D^b(Coh(\CC^*\times\CC[2]))$. On the A-side, we have an A-model with target $T^*S^1$ tensored with an A-type 2d gauge theory with gauge group $U(1)$. One could guess that the corresponding category of branes is a $U(1)$-equivariant version of the Fukaya-Floer category of $T^*S^1$. More generally, one could guess that the category of branes in an A-model with target $T^*X$  tensored with the A-type 2d $U(1)$ gauge theory is a $U(1)$-equivariant version of the Fukaya-Floer category of $T^*X$.  It is not clear to us how to define such an equivariant Fukaya-Floer category mathematically. Given the results of  \cite{ZN,Nadler}, a natural guess is the equivariant constructible derived category of sheaves on $X$. As a check, note that when $X$ is a point, the $U(1)$-equivariant constructible derived category is equivalent to $D^b(Coh(\CC[2]))$ \cite{BL}. As mentioned above and explained in appendix F, this is indeed the category of branes for the A-type 2d gauge theory. The monoidal structure seems to be the standard one (derived tensor product). On the B-side, on the other hand, the monoidal structure is a combination of the tensor product of coherent sheaves on $\CC[2]$ and the convolution product on $\CC^*$. 

\subsubsection{The DN condition}

Next consider the boundary condition on the A-side which is a combination of the Dirichlet condition in the gauge sector and the Neumann condition for $A_4$ in the matter sector. It is dual to the Dirichlet condition for the B-type gauge sector and a boundary condition for the RW model with target $T^*[2]\CC^*$ which sets $\sigma=0$ and leaves the complex scalar $\tau=A_4+i\phi_4$ free to fluctuate. 

On the B-side reduction on an interval  gives a B-model with target $\CC^*\times\CC^*$, therefore the category of boundary line operators is $D^b(Coh(\CC^*\times\CC^*))$. On the A-side reduction gives an A-model with target $T^*U(1)\times T^*U(1)$, therefore the category of boundary line operators is the Fukaya-Floer category. The two categories are equivalent by the usual 2d mirror symmetry. The monoidal structure is easiest to determine on the B-side. It is neither the derived tensor product, nor the convolution, but a combination of both. This happens because the two copies of $\CC^*$ have a very different origin: one of them arises from a 3d B-type gauge theory, and the other one arises from the Rozansky-Witten model with target $T^*[2]\CC^*$.

\subsubsection{The ND condition}

Finally we consider the boundary condition on the A-side which is a combination of the Neumann condition in the gauge sector and the Dirichlet condition for $A_4$. This is the case which corresponds to the Gukov-Witten surface operator at $t=1$. Indeed, the Dirichlet conditions for $A_4,\phi_4$ and $\phi_3$ mean that the holonomy of $A$ is fixed, while the 1-form $\phi$ has a singularity of the form
$$
\beta \frac{dr}{r}-\gamma d\theta,
$$
where $-\gamma$ is the boundary value of $\phi_4$ and $\beta$ is the boundary value of $\phi_3$. The boundary value of $A_4$ is the Gukov-Witten parameter $\alpha$. The Neumann condition in the gauge sector also depends on the boundary theta-angle which corresponds to the Gukov-Witten parameter $\eta$. As explained above, the boundary values of $A_4$ and $\phi_3$ are actually irrelevant. This agrees with the results of \cite{GW}, where it is shown that at $t=1$ the parameters $\alpha$ and $\beta$ are irrelevant. Thus the true parameter space of the surface operator on the A-side is $\CC^*$.

Electric-magnetic duality maps the DD condition to the Neumann condition for the B-type gauge theory and the boundary condition in the RW model which fixes $\tau$ and leaves $\sigma$ free to fluctuate. The latter boundary condition depends on the boundary value of the field $\tau=A_4+i\phi_4$. From the 4d viewpoint this boundary value encodes the Gukov-Witten parameters $\alpha$ and $\gamma$. These are the relevant parameters at $t=i$, as explained in \cite{GW}. The Neumann boundary condition in the B-type gauge theory also has two parameters (the boundary value of $\phi_3$ and the boundary theta-angle) which correspond to the Gukov-Witten parameters $\beta$ and $\eta$. But as explained above and from a different viewpoint in \cite{GW}, these parameters are irrelevant at $t=i$.

Let us compare the categories of 3d boundary line operators, which from the 4d viewpoint are interpreted as categories of line operators sitting on Gukov-Witten surface operators. On the B-side reduction on an interval gives a B-model with target $\CC[2]$ tensored with a B-type 2d gauge theory, therefore the category of boundary line operators is $D^b_{\CC^*}(Coh(\CC[2]))$. On the A-side reduction on an interval gives an A-type 2d gauge theory tensored with a modified A-model with target $T^*\RR$. Its category of branes is a modification of the category of boundary conditions for the A-type 2d gauge theory where the space of boundary degrees of freedom has additional integer grading coming from the winding of the periodic scalar $A_4$, and morphisms are required to have degree zero with respect to it. Since branes in the A-type 2d gauge theory can be identified with objects of $D^b(Coh(\CC[2]))$, the category of boundary conditions in the combined system is equivalent to $D^b_{\CC^*}(Coh(\CC[2]))$, in agreement with what we got on the B-side.

\subsection{A proposal for the 2-category of surface operators at $t=1$}

By analogy with the Rozansky-Witten model, one may conjecture that the 2-category of surface operators at $t=0$ can be described in terms of module categories over the monoidal category of boundary line operators for the distinguished boundary condition (the NN condition). We have argued above that this monoidal category is the $U(1)$-equivariant constructible derived category of $S^1$, where the $U(1)$ action on $S^1$ is trivial. It is probably better to think about it as a sheaf of $U(1)$-equivariant monoidal DG-categories over $S^1$. To each surface operator we may associate a sheaf of $U(1)$-equivariant module categories over this sheaf of $U(1)$-equivariant monoidal categories, and we conjecture that this map is an equivalence of 2-categories. Gukov-Witten-type operators correspond to skyscraper sheaves on $S^1$. 

Electric-magnetic duality then implies that there is an equivalence between this 2-category and the 2-category of coherent $\CC^*$-equivariant derived categorical sheaves over $\CC^*$.

\section{Surface operators at $t=0$}

\subsection{Reduction to 3d}

The 3d theory again decomposes into the gauge and matter sectors. Let us start with the gauge sector. The bosonic fields are a gauge field $A$, a periodic scalar $A_4$, and a complex scalar $\sigma$. The fermionic fields are two 0-forms $\eta$ and $\psi_4$, a 1-form $\psi$ and a 2-form $\chi_+$. Thus subscript $+$ indicates that $\chi_+$ originates from the self-dual part of the 2-form $\chi$ in four dimensions.The BRST transformations are
\begin{align*}
\delta A &=i\psi,\\
\delta A_4 &=i\psi_4,\\
\delta\psi &=d\sigma,\\
\delta\psi_4 &=0,\\
\delta\sigma &=0,\\
\delta\bsigma &=i\eta,\\
\delta\eta &=0,\\
\delta\chi_+ &= F+\star dA_4
\end{align*}
The field content and BRST transformations are the same as in the A-type 3d gauge theory, the main difference being that the bosonic scalar $A_4$ is periodic. The action of the gauge sector contains, apart from a BRST-exact term, a topological term
\begin{equation}\label{topaction}
S_{top}=-\frac{2\pi}{e^2}\int_{M_3} F\wedge dA_4
\end{equation}
Note that it is the periodicity of $A_4$ that makes this topological term nontrivial in general. 

The above topological term term comes from the dimensional reduction of a topological term in 4d
$$
-\frac{1}{2e^2}\int F\wedge F.
$$
Here we assumed that the 4d theta-angle vanishes and that the coordinate $x^4$ has period $2\pi$.  

In the matter sector the only bosonic fields are a 0-form $\phi_4$ and a 1-form $\phi$. The fermionic fields are a pair of 0-forms $\teta$ and $\tpsi_4$, a 1-form $\tpsi$, and a 2-form $\chi_-$ which arises from the anti-self-dual part of the 2-form $\chi$ in four dimensions. The matter content and BRST transformations are the same as for the $t=1$ matter sector, except that the periodic scalar $A_4$ is replaced with a non-periodic scalar $\phi_4$. The matter action is BRST-exact.

\subsection{The gauge sector}

As for $t=1$, we may consider either Dirichlet or Neumann conditions for the gauge field, and then BRST-invariance determines the rest. The category of boundary line operators is determined by compactifying the theory on an interval with the appropriate boundary conditions and analyzing branes in the resulting 2d TFT.

In the Neumann case the effective 2d TFT is the A-type 2d gauge theory, just as for $t=1$. As explained above, its category of boundary conditions is equivalent to $D^b(Coh(\CC[2]))$. 

In the Dirichlet case the effective 2d TFT is a topological sigma-model with two bosonic fields, $A_4$ and the holonomy of the 3d gauge field $A$ along the interval. Both are periodic scalars, so the target of the sigma-model is $T^2$.  The BPS equations reduce to a holomorphic instanton equation
$$
dA_3+\stard dA_4=0,
$$
which means that we are dealing with an A-model with target $T^2$. Its category of branes is the Fukaya-Floer category of $T^2$, which is fairly nontrivial (and by mirror symmetry equivalent to the bounded derived category of coherent sheaves on an elliptic curve). The A-model depends on the symplectic form on $T^2$ which can be read off the topological piece of the action (\ref{topaction}). Setting $A_1$ and $A_2$ to zero and reducing on an interval of length $2\pi$ it becomes

$$
\frac{-4\pi^2 }{e^2}\int_{M_2} dA_3\wedge dA_4
$$
We may regard this expression as an integral of the pull-back of a symplectic 2-form 
$$
\frac{4\pi^2}{e^2} dx\wedge dy. 
$$
on the 2-torus with periodic coordinates $x,y$, both with period one. The symplectic area of this 2-torus is $4\pi^2/e^2$. 

We do not know how to describe the monoidal structure on this category arising from the fusion of boundary line operators.

\subsection{The matter sector}

As for $t=1$, we may consider either the Dirichlet or Neumann conditions for the scalars $\phi_3$ and $\phi_4$ (BRST-invariance requires them to be of the same type). In the Dirichlet case reduction on an interval gives the modified A-model whose only bosonic field is a real 1-form $\phi$ in two dimensions. As discussed above, it category of branes is the same as for a trivial TFT, i.e. it is equivalent to $D^b(Coh(\bul))$. Unlike in the $t=1$ case, there are no ``winding sectors,'' since the scalars $\phi_3$ and $\phi_4$ are not periodic. So the category of boundary line operators in this case is $D^b(Coh(\bul))$, with its standard monoidal structure.

If $\phi_3$ and $\phi_4$ satisfy the Neumann condition, then the restriction of the 1-form $\phi$ to the 2d boundary must vanish. Reducing on an interval, we get an A-model whose only bosonic fields are $\phi_3$ and $\phi_4$, namely an A-model with target $T^*\RR$. Its category of branes is the Fukaya-Floer category of $T^*\RR$. Since this should be thought as the category of boundary line operators in a 3d TFT, it should have a monoidal structure. Since the only difference compared to the $t=1$ matter sector is the noncompactness of $\phi_4$, we expect that after we apply the equivalence of \cite{Nadler}, this monoidal structure becomes the standard monoidal structure on the constructible derived category of $\RR$. 

\subsection{Putting the sectors together}

\subsubsection{The DD condition}

The DD boundary condition corresponds to a surface operator such that the 1-form $\phi$ has a fixed singularity of the form
$$
\beta \frac{dr}{r}-\gamma d\theta,
$$
while the holonomy of the gauge field $A$ is allowed to fluctuate, and the scalar field $\sigma$ vanishes at the insertion surface. To define such an operator properly, one has to excise a tubular neighborhood of the insertion surface and impose suitable conditions on the newly created boundary. 

Since the matter sector in the Dirichlet case does not have interesting boundary conditions, the category of boundary line operators is the same as in the gauge sector, i.e. the Fukaya-Floer category of $T^2$ with the symplectic area ${\mathfrak S}=4\pi^2/e^2$. From the 4d viewpoint, this is the category of line operators on the surface operator. 

Electric-magnetic duality maps the DD condition to itself. Indeed, it does not affect the matter sector, while in the gauge sector it maps the periodic scalar $A_4$ into a gauge field and maps the gauge field to a periodic scalar. Since in the DD case $A_4$ satisifies the Neumann condition, the dual gauge field satisfies the Dirichlet condition. Contrariwise, the Dirichlet condition for the gauge field is mapped by duality to the Dirichlet condition for the new periodic scalar. The only effect of duality is to replace $e^2$ with $4\pi^2/e^2$. Therefore the symplectic area of the $T^2$ is also inverted:
$$
{\mathfrak S}\mapsto {\mathfrak S}'=\frac{4\pi^2}{{\mathfrak S}}
$$
The Fukaya-Floer categories of two tori whose symplectic areas are related as above are equivalent by the usual T-duality. Moreover, we expect that the monoidal structure (which we have not determined!) is preserved by T-duality.

\subsubsection{The NN condition}

The NN condition corresponds to the surface operator such that $A$ has a fixed singularity of the form
$$
\alpha d\theta,
$$
while the singularity for the 1-form $\phi$ is allowed to fluctuate. To define such a surface operator properly, one has to impose suitable conditions on a boundary of a tubular neighborhood of the insertion surface. 

Upon reduction on an interval with NN boundary conditions on both ends, we get a 2d TFT which is a product of an A-type 2d gauge theory and an A-model with target $T^*\RR$. 
Its category of branes is an equivariant version of the Fukaya-Floer category of $T^*\RR$. It was conjectured above that it is equivalent to the equivariant  constructible derived category of $\RR$, with the standard monoidal structure (derived tensor product).

Electric-magnetic duality maps the NN condition to itself, for the same reason as in the DD case. It acts trivially on the category of line operators, because the bosonic fields which survive the reduction on an interval (that is, $\sigma$, $\phi_3$ and $\phi_4$) are not involved in the duality. 

\subsubsection{The DN condition}

The DN condition corresponds to a surface operator such that both $A$ and $\phi$ are allowed to have fluctuating singularities, while $\sigma$ has to vanish at the surface operator. Upon reduction on an interval with DN boundary conditions on both ends, we get a product of an A-model with target $T^2$ and an A-model with target $T^*\RR$. Its category of branes is the Fukaya-Floer category of $T^2\times T^*\RR$. Electric-magnetic duality maps the DN condition to itself. Its action on the category of line operators amounts to a T-duality on $T^2$ (duality acts trivially on the matter sector). The monoidal structure (which we have not determined) must be preserved by T-duality.

\subsubsection{The ND condition}

This case corresponds to the Gukov-Witten surface operator where the holonomy of $A$ is fixed, and the 1-form $\phi$ has a fixed singularity of the form
$$
\beta \frac{dr}{r}-\gamma d\theta.
$$

Reduction on an interval with ND boundary conditions gives a 2d TFT which is a product of an A-type 2d gauge theory and a modified A-model whose only bosonic field is a real 1-form. Since there are no interesting boundary conditions in the latter theory, the category of boundary conditions in this case is the same as in the former theory. That is, it is the $U(1)$-equivariant constructible derived category of sheaves over a point, or equivalently $D^b(Coh(\CC[2]))$ \cite{BL}. This is therefore the category of line operators sitting on the Gukov-Witten surface operator. The monoidal structure is the standard one (derived tensor product).

In particular, since the trivial surface operator is a special case of the Gukov-Witten surface operator, we conclude that the category of bulk line operators in the GL-twisted theory at $t=0$ is $D^b(Coh(\CC[2]))$. In 4d terms, this can be interpreted as saying that all bulk line operators can be constructed by taking a sum of several copies of the trivial line operator and deforming it using the descendants of the BRST-invariant field $\sigma$ and its powers. This agrees with the results of \cite{Kap:qGL}, where it was argued that neither Wilson nor 't Hooft line operators are allowed at $t=0$.

Electric-magnetic duality maps the ND condition to itself. It acts trivially on the category of line operators since the field $\sigma$ is not involved in the duality.

\section{Conclusions}

We have seen that GL-twisted gauge theory has a large number of surface operators other than the Gukov-Witten surface operators. These surface operators can be organized into a 2-category, and in the case $t=i$ we also proposed a description of this 2-category in terms of module categories. For $G=U(1)$ we proposed a similar description at $t=1$. It would be very interesting to find a physically-motivated\footnote{It was proposed by D. Gaitsgory that for $t\neq \pm i$ this 2-category can be described in terms categories with a D-module action of the loop group of $G$, but it is not clear how this proposal is related to the physical picture.} description of the 2-category of surface operators for all $t$ and $G$. Montonen-Olive duality implies that the 2-category of surface operators at $t=i$ in a theory with gauge group $G$ is equivalent to the 2-category of surface operators at $t=1$ in a theory with gauge group $\LG$. Moreover, these 2-categories both have braided monoidal structure, and the equivalence must be compatible with them. The usual statement about the equivalence of the categories of Wilson and 't Hooft line operators in the two theories follows from this. Indeed, bulk line operators can be regarded as endomorphisms of the trivial surface operator, and so must be equivalent (as tensor categories). From the mathematical viewpoint, the statement about the equivalence of braided monoidal 2-categories can be regarded as a 2-categorification of the geometric Satake correspondence.\footnote{The original version of the geometric Satake correspondence is due to Lusztig \cite{Lusztig} and can be regarded as a statement about the K-theory of the category of bulk line operators in the 4d TFT. That is, it is a statement about the commutative algebra which the 4d TFT attaches to $S^2\times S^1$. A way to categorify it to replace $S^2\times S^1$ with $S^2$; this corresponds to studying the symmetric monoidal category of line operators in the 4d TFT. This version of the geometric Satake correspondence has been proved by Ginzburg \cite{Ginz} and Mirkovic and Vilonen \cite{MV}. Alternatively, we obtain a 2-categorification of the geometric Satake correspondence by replacing $S^2\times S^1$ with $S^1$. Physically this corresponds to studying the braided monoidal category of surface operators in the 4d TFT.}

The 2-category of surface operators at $t=0$ is a natural setting for studying local quantum geometric Langlands. In this case Montonen-Olive duality should give a nontrivial equivalence of braided monoidal 2-categories of surface operators in theories whose gauge couplings are inversely related. Already in the abelian case we saw that this equivalence is fairly nontrivial and reduces to T-duality in some special cases. 

As for the global quantum geometric Langlands, it seems natural to study the 2d TFT obtained by compactifying the 4d TFT at $t=0$ on a Riemann surface $C$ with an insertion of a surface operator of type DN or DD. This means that we cut a hole in $C$ and impose a boundary condition which allows the holonomy of $A$ along the boundary of the hole to be arbitrary. The resulting effective 2d TFT will have a category of branes which is a module category over the monoidal category of surface line operators of type DN or DD. The Montonen-Olive duality implies that replacing the group $G$ by its Langlands dual and inverting the gauge coupling gives rise to an equivalence of monoidal categories of surface line operators and a compatible equivalence of the categories of branes.

We described the category of bulk line operators in the 4d theory at $t=i$, for a general gauge group. We found that it is equivalent to the equivariant derived category of coherent sheaves on the Lie algebra of the gauge group, with the linear coordinates on the Lie algebra sitting in cohomological degree $2$. This is a much larger category than one might naively expect based on special examples such as Wilson line operators. 

Finally, we showed that for $t=0$ and abelian gauge group the category of bulk line operators is fairly small (equivalent to $D^b(Coh(\CC[2]))$), and that electric-magnetic duality acts trivially on it. This agrees with \cite{Kap:qGL}, where it was shown that the $t=0$ theory does not admit either Wilson or 't Hooft line operators. It seems plausible that for a general gauge group electric-magnetic duality acts trivially on the category of bulk line operators at $t=0$.

\appendix

\section{B-type topological gauge theory in 2d}

In this appendix we discuss a topological gauge theory  in 2d which can be obtained by twisting $N=(2,2)$ supersymmetric gauge theory by means of a $U(1)_A$ current. This theory is a 2d analog of the GL-twisted theory at $t=i$.  

The fields of the B-type 2d gauge theory are a connection $A$ on a principal $G$-bundle $\cP$ over an oriented 2-manifold $M_2$, a 1-form $\phi$ with values in $\Ad(\cP)$, a fermionic 0-form $\beta$ with values in $\Ad(\cP)$, a fermionic 1-form $\lambda$ with values in $\Ad(\cP)$, and a fermionic 2-form $\zeta$ with values in $\Ad(\cP)$. The BRST transformations are
\begin{align*}
\delta A &=\lambda,\\
\delta \phi &=i\lambda,\\
\delta\lambda &=0,\\
\delta\beta &=i d_A^\stard \phi,\\
\delta\zeta &=-i \cF
\end{align*}
Here $\cF$ is the curvature of the complex connection $\cA=A+i\phi$. We will denote the covariant derivative with respect to $\cA$ by $d_\cA$, the covariant derivative with respect to $\bar\cA=A-i\phi$ by $d_\cA$, and the curvature of $\bar\cA$ by $\bar\cF$. The differential operator $d_A^\stard$ is $\stard d_A\stard$, where $\stard$ is the 2d Hodge star.

The theory has a $U(1)$ ghost number symmetry with respect to which the fields $A,\phi$ are neutral, the field $\lambda$ has charge $1$, and the fields $\beta,\zeta$ have charge $-1$. 

The BRST transformation satisfy $\delta^2=0$ on all fields except $\beta$:
$$
\delta^2\beta =-d_\cA^\stard\lambda.
$$
If one uses the fermionic equation of motion $d_\cA^\stard\lambda=0$, then the BRST transformations are nilpotent on-shell. It is more convenient to have $\delta^2=0$ off-shell, so we introduce an auxiliary 0-form $P$ and define
$$
\delta\beta=iP,\quad \delta P=0.
$$
When constructing an action, we need to ensure that the equation of motion for $P$ sets $P=d_A^\stard\phi$. A suitable action is BRST-exact:
$$
S=-\frac{1}{2e^2}\delta \int_{M_2} \Tr\,\left(i\zeta\wedge \stard\bar\cF+i\beta\wedge \stard  (P-2 d_A^\stard\phi)\right).
$$
The coupling constant $e^2$ enters only as the coefficient of a BRST-exact term, therefore the topological correlators do not depend on it, and the semiclassical approximation is exact. The topological nature of the theory is also apparent, since the metric enters only through BRST-exact terms. 

Usually local observables are defined as BRST-invariant and gauge-invariant scalar functions of fields, modulo BRST transformations. In the present case, there are no nontrivial local observables of this kind. However, there are nontrivial BRST-invariant local disorder operators which are defined by allowing certain singularities in the fields. For example, one can require the connection $\cA$ to have a nontrivial holonomy around the insertion point. Such local operators are analogous to Gukov-Witten surface operators in 4d gauge theory. More systematically, to determine what kind of local operators are allowed one can reduce the 2d gauge theory theory on a circle and study the space of the states of the resulting 1d TFT. In the present case, this 1d TFT is a gauged sigma-model with target $G_\CC$. From the 2d viewpoint, the target space parameterizes the holonomy of $\cA$. BRST-invariant wave-functions are holomorphic functions on $G_\CC$ invariant with respect to conjugation, i.e. characters of $G_\CC$. More generally, one may consider non-normalizable wavefunctions, such as delta-functions supported  on closed $G_\CC$-invariant complex submanifolds of $G_\CC$. For example, the identity operator can be thought of as a delta-function supported at the identity element, while Gukov-Witten-type local operators are delta-functions supported on closed conjugacy classes in $G_\CC$. 

There are also BRST-invariant and gauge-invariant line observables, the most obvious of which are Wilson line operators for the complex BRST-invariant connection $\cA$. To define them, one needs to pick a finite-dimensional graded representation $V$ of $G$ and consider the holonomy of $\cA$ in the representation $V$.

The category of branes for this 2d TFT is the category of finite-dimensional graded representations of $G$. To see this, consider the Neumann boundary condition for the gauge field, that is, leave the restriction of $\cA$ to the boundary free and require the restriction of $\stard\phi$ to vanish. BRST-invariance then requires $\zeta$ and the restriction of $\stard\lambda$ to vanish on the boundary. Since the gauge field $\cA$ on the boundary is unconstrained and BRST-invariant, we may couple to it an arbitrary finite-dimensional graded representation $V$ of $G$. That is, we may include into the path-integral the holonomy of $\cA$ in the representation $V$. Thus boundary conditions are naturally labeled by representations of $G$. Given any two irreducible representations $V_1$ and $V_2$ one can form a junction between them only if $V_1$ and $V_2$ are isomorphic (because there are no nontrivial  BRST-invariant local operators on the Neumann boundary). Further, if $V_1\simeq V_2$, the space of morphisms between them is $\Hom_G(V_1,V_2)$ (for the same reason). 

\section{The gauged B-model}

In this appendix we describe how to couple a B-type 2d gauge theory to a B-model. We show that the category of branes for the resulting 2d TFT is closely related to the equivariant derived category of coherent sheaves. 

Let $X$ be a Calabi-Yau manifold (i.e. a K\"ahler manifold with a holomorphic volume form) which admits a $G$-action which preserves the K\"ahler structure and the holomorphic volume form. The infinitesimal action of $G$ is described by a holomorphic vector field $V^I$ with values in $\frg^*$ (the dual of the Lie algebra of $G$).  The fields of the usual B-model are a map $\sigma: M_2\ra X$, a fermionic 0-form $\eta\in \sigma^*\overline{TX}$, a fermionic 0-form $\theta\in\sigma^*T^*X$, and a fermionic 1-form $\rho\in T^*M_2\otimes \sigma^*TX$. Consider the following BRST transformations:
\begin{align*}
\delta\sigma^I&=0,\\
\delta \sigma^\bI&=\eta^\bI,\\
\delta\eta^\bI &=0,\\
\delta\theta_I &=0,\\
\delta\rho^I &=d\sigma^I+V^I(\cA)=\cD\sigma^I.
\end{align*}
This is a covariantized version of the usual B-model BRST transformations. The appearance of the covariant derivative $\cD\sigma^I$ means that $\sigma$ is now interpreted as a section of a fiber bundle over $M_2$ with typical fiber $X$ which is associated to a principal $G$-bundle $\cP$ over $M_2$. Since the connection $\cA$ is BRST-invariant, these BRST transformations still satisfy $\delta^2=0$.

To construct a BRST-invariant action we take the usual action of the B-model and covariantize all derivatives. The covariantized action is not BRST-invariant, but this can be corrected for by adding a new term proportional to $\theta_I V^I(\zeta)$, where $\zeta$ is the fermionic $\Ad(\cP)$-valued 2-form which is part of the B-type 2d gauge theory. The full matter action is
$$
S=\int_{M_2} \delta\left( g_{I\bJ} \rho^I \wedge \stard {\bar\cD\sigma}^\bJ\right)+\int_{M_2}\left(-i\theta_I V^I(\zeta)+\theta_I \cD\rho^I+\frac12 R^I_{JK{\bar L}} \theta_I\rho^J\rho^K\eta^{\bar L}\right).
$$
Here $g_{I\bJ}$ is the K\"ahler metric, $R$ is its curvature tensor, and the covariant derivative of $\rho$ includes both the Levi-Civita connection and the gauge connection:
$$
\cD\rho^I=d\rho^I+\Gamma^I_{JK} d\sigma^J\rho^K+\nabla_J V^I(\cA)\rho^J,\quad \nabla_J V^I=\partial_J V^I+\Gamma^I_{JK}V^K.
$$
The covariant derivative $\bar\cD\sigma^\bJ$ is defined so as to make the bosonic part of the action positive-definite:
$$
\bar\cD\sigma^\bJ=d\sigma^\bJ-V^\bJ(\bar\cA),\quad \bar\cA=A-i\phi=-\cA^\dagger.
$$

Since the category of branes for the B-model with target $X$ is $D^b(Coh(X))$, a natural guess for the category of branes for the gauged B-model is $D^b_{G_\CC}(Coh(X))$. 
We will now describe a construction of the boundary action corresponding to an equivariant complex of holomorphic vector bundles on $X$. Let $E$ be a graded complex vector bundle over $X$ with a holomorphic structure $\bpartial^E: E\ra E\otimes\Obul(X)$, $(\bpartial^E)^2=0$, and a holomorphic degree-1 endomorphism $T:E\ra E$, $\bpartial^E T=0$ satisfying $T^2=0$.  To write down a concrete boundary action we will assume that we are also given a Hermitian metric on each graded component of $E$, so that $\bpartial^E$ gives rise to a connection $\nabla^E$ on $E$. We will denote the corresponding connection 1-form by $\omega$ and its curvature by $F^E$. We assume that we are given a lift of the $G$-action on $X$ to a $G$-action on the total space of $E$ which is fiberwise-linear and compatible with $\bpartial^E$, $T$, and the Hermitian metric. Infinitesimally, the Lie algebra $\frg$ acts on a section $s$ of $E$ as follows:
$$
(f,s)\mapsto f(s)=V^I(f)\nabla^E_I s+V^\bI(f)\nabla^E_\bI s+R(f)s,\quad f\in\frg .
$$
Here $\nabla^E=d+\omega$, and $R$ is a degree-0 bundle morphism $R:E\ra E\otimes \frg^*$. The condition that the $G$-action commutes with $\nabla^E$ implies
$$
\nabla^E R=\iota_V F^E.
$$
The condition that the $G$-action commutes with $T$ implies
$$
V^I\nabla^E_I T+[R,T]=0.
$$
Consider now the following field-dependent connection 1-form on the pull-back bundle $\sigma^*E$:
$$
\cN=\omega_I d\sigma^I+\omega_\bI d\sigma^\bI-R(\cA)+\rho^I\eta^\bJ F^E_{I\bJ} +\rho^I\nabla^E_I T.
$$
With some work one can check that its BRST variation satisfies
$$
\delta \cN=d(\omega_\bI\eta^\bI+T)+[\cN,\omega_\bI\eta^\bI+T].
$$
Therefore the supertrace of its holonomy is BRST-invariant and can be used as a boundary weight factor in the path-integral associated. By definition the boundary action is minus the logarithm of the boundary weight factor.

Let us consider a ghost-number zero boundary observable $\cO$ in the presence of a such a weight factor. It is an element of $\End(E)$ depending on the fields $\sigma,\eta$ and of total degree zero. More invariantly, we may think of it as a section of $\End(E)\otimes \Obul(X)$. The BRST-variation of the boundary weight factor in the presence of $\cO$ is proportional to
$$
\eta^\bI \nabla^E_\bI \cO+[T,\cO].
$$
Hence BRST-invariant boundary observables are sections of $\End(E)\otimes\Obul(X)$ which are annihilated by $\bpartial^E$ and commute with $T$. Further, a BRST-invariant $\cO$ it is gauge-invariant iff it satisfies
$$
V^I\nabla^E_I \cO+[R,\cO]=0
$$
Together these conditions mean that $\cO$ represents an endomorphism of the equivariant complex $(E,\bpartial^E,T)$ regarded as an object of $D^b_{G_\CC}(Coh(X))$. It is also easy to see that such an observable $\cO$ is a BRST-variation of a gauge-invariant observable iff it is homotopic to zero.  In some cases this implies that the category of branes in the gauged B-model of the kind we have constructed is equivalent to $D^b_{G_\CC}(Coh(X))$. This happens if any $G$-equivariant coherent sheaf on $X$ has a $G$-equivariant resolution by $G$-equivariant holomorphic vector bundles. Such an $X$ is said to have a $G$-resolution property. An example of such $X$ is $\CC^n$ with a linear action of $G$, or more generally a smooth affine variety with an affine action of $G$. Note that for a general complex manifold $X$ the resolution property may fail even if $G$ is trivial. But for trivial $G$ the cure is known: one has to replace complexes of holomorphic vector bundles with more general DG-modules over the Dolbeault DG-algebra of $X$ \cite{Block}. These more general DG-modules also arise naturally from the physical viewpoint \cite{Bergman,KRS}.  We expect that for any complex Lie group $G$ with a complex-analytic action on $X$ a $G$-equivariant coherent sheaf on $X$ has a $G$-equivariant resolution by these more general DG-modules. This would imply that the category of B-branes for the gauged B-model is equivalent to $D^b_{G_\CC}(Coh(X))$.

\section{B-type topological gauge theory in 3d}

Consider Euclidean $N=4$ $d=3$ SYM theory.  In addition to the $SU(2)_{E}$ rotational symmetry, this theory has an $SU(2)_{\mathcal{R}} \times SU(2)_{N}$ $\mathcal{R}$-symmetry.  This affords us two distinct topological twists on a generic 3-manifold.  Twisting by $SU(2)_{\mathcal{R}}$ gives the dimensional reduction of the Donaldson-Witten theory, which resembles the A-model.  Twisting by the $SU(2)_{N}$ symmetry leads to a topological gauge theory resembling the B-model.  We therefore refer to these topological gauge theory as A-type and B-type, respectively. 

In this appendix we describe the B-type 3d gauge theory which was first constructed by Blau and Thompson \cite{BT}.  The field content consists of a gauge field $A$, a bosonic 1-form field $\phi$, a fermionic 2-form field $\zeta$, a fermionic 1-form field $\lambda$, and two fermionic scalars $\rho$, $\tilde{\rho}$.  Furthermore, it is convenient to introduce an auxiliary scalar field $P$.  All fields are in the adjoint representation of the gauge group $G$. The action for this theory is
\begin{equation} \label{eq:SOGL_App1_BT_action}
	\begin{split}
		\tilde{S} & = - \frac{1}{2e^{2}} \int_{M_{3}} \Tr \Big( \mathcal{F} \wedge \star \bar{\mathcal{F}} - P \wedge \star \big( P - 2 d^\star_{A} \phi \big) - 2i \zeta \wedge \star d_{\bar{\mathcal{A}}} \lambda - 2 \tilde{\rho} \wedge \star d^\star_{\mathcal{A}} \lambda - 2 e^{2} \rho \wedge d_{\mathcal{A}} \zeta \Big)
	\end{split}
\end{equation}
where $\mathcal{A}$, $\bar{\mathcal{A}}$ are the complexified connections $A \pm i \phi$ and $\mathcal{F}$, $\bar{\mathcal{F}}$ are the corresponding field strengths.  The BRST variation of the fields are
\begin{equation} \label{eq:SOGL_App1_BT_variations}
	\begin{aligned}
		\delta A & = \lambda, \\
		\delta \phi & = i \lambda, \\
		\delta \lambda & = 0, \\
		\delta \zeta & = - i \mathcal{F}, \\
		\delta \rho & = 0, \\
		\delta \tilde{\rho} & = i P, \\ 
		\delta P & = 0.
	\end{aligned}
\end{equation}
There is also a ghost number symmetry with respect to which $\rho$ and $\lambda$ have charge $1$ and $\zeta$ and $\tilde\rho$ have charge $-1$.
The action is BRST exact up to a metric independent term:
\begin{equation}
	\begin{split}
		\tilde{S} & = - \frac{1}{2e^{2}} \delta \int_{M_{3}} \Tr \Big( i \zeta \wedge \star \bar{\mathcal{F}} + i \tilde{\rho} \wedge \star \big( P - 2 d^\star_{A} \phi \big) \Big) + \int_{M_{3}} \Tr \big( \rho \wedge d_{\mathcal{A}} \zeta \big) . 
	\end{split}
\end{equation}
The equation of motion for $P$ reads $P=d^\star_A\phi$; if we substitute this value into the BRST transformations, the BRST operator is only nilpotent modulo the fermionic equations of motion.

It is more natural to regard the fermionic 0-form $\rho$ as taking values in $\frg^*$ rather than $\frg$, because then the non-BRST-exact piece in the action takes the form
$$
\int_{M_3} \rho_a \wedge d_\cA\zeta^a,
$$
which is manifestly  independent of the choice of metric on $\frg$.\footnote{For a simple Lie algebra $\frg$ there is a canonical identification of $\frg$ and $\frg^*$ by means of the Killing form, but if $\frg$ has an abelian subalgebra, the metric is not uniquely determined.}

Local observables in this topological gauge theory are gauge invariant functions of $\rho$, which correspond to elements in the exterior algebra $\Lambda^\bul(\mathfrak{g})$ invariant with respect to the adjoint action. Unlike in the 2d case, there are no disorder local operators.

The simplest line operators can be constructed as Wilson lines for the BRST-invariant connection $\cA=A+i\phi$. Such operators are labeled by finite-dimensional representations of $G$. More generally, we may consider coupling a 1d TFT living on the line to the 3d gauge theory. The space of states for this 1d TFT is a $\ZZ$-graded vector space $V$.  Endomorphisms of $V$ are naturally graded as well.  Let us denote the degree-1 endomorphism that generates the BRST symmetry in this theory as $T(\Phi) \in \End(V)$ and the degree-0 endomorphisms that generate the gauge symmetry as $R_{a}(\Phi) \in \End(V)$, where $\Phi$ represents the fields in the topological gauge theory.  
Since $\rho$ is the only BRST-invariant 0-form in the 3d gauge theory, it is sufficient to assume that $T$ is a function of $\rho$ alone. It also must be nilpotent:
\begin{align}
 	T(\Phi) & = T(\rho), \\
	T^{2} & = 0.
\end{align}
Since the gauge symmetry preserves the grading of $V$, the generators of the gauge symmetry $R_{a}$ may be assumed to be $\rho$-independent:
\begin{align}
 	R_{a}(\Phi) & = R_{a}.
\end{align}
Since the gauge symmetry $\delta_g$ and BRST symmetry $\delta$ commute in the 3d gauge theory, $T$ and $R_{a}$ must satisfy the following relation:
\begin{equation}\label{Tcovariance}
	\begin{split}
		0 & = \big[ \delta_{g}(f), \delta \big] \\
		& =  \big[ f, \rho \big]^{a} \frac{\partial T}{\partial \rho^{a}} + \big[ f^{a} R_{a}, T \big]
	\end{split}
\end{equation}
where $f \in \mathfrak{g}$. To construct  the line observable associated to the triple $(V,T,R)$, we apply the descent procedure to $T$ to get a connection 1-form on the graded vector bundle with fiber $V$. By definition, the descendant connection $\cN$ is defined by the  equation
$$
\delta \cN=dT+[\cN,T].
$$
Using the relation (\ref{Tcovariance}) and the fermionic equations of motion, we find 
\begin{equation}
	\mathcal{N} = \frac{i}{2 e^{2}} \star \bar{\mathcal{F}}^a \frac{\partial T}{\partial \rho^{a}} + \mathcal{A}^{a} R_{a}.
\end{equation}
The supertrace of the holonomy of $\mathcal{N}$ along a curve $\gamma$ in $M_{3}$ is therefore a BRST invariant, gauge invariant loop operator in the topological gauge theory. The holonomy itself defines a line operator.

Line operators in any 3d TFT form a braided monoidal category. The subcategory formed by line operators described above is the $G$-equivariant derived category of DG-modules over the DG-algebra $\Lambda^\bul(\frg)$ (with zero differential). To see this, consider a local operator inserted at the junction of two Wilson lines corresponding to the triples $(V_1,T_1,R_1)$ and $(V_2,T_2,R_2)$. Since we are looking for BRST-invariant operators, one may assume that it is a function $\cO$ of $\rho$ valued in $\Hom_\CC(V_1,V_2)$, or in other words an element of $\Hom_\CC(V_1,V_2\otimes \Lambda^\bul(\frg))$. The BRST-operator acts on $\cO$ by
$$
\delta \cO =T_2 \cO \pm \cO T_1
$$
where the sign is plus or minus depending on whether the total degree of $\cO$ is odd or even. Gauge transformations act on $\cO$ in the obvious way and commute with the BRST operator. The space of morphisms between the line operators is the cohomology of $\delta$ on the $G$-invariant part of $\Hom_\CC(V_1,V_2\otimes \Lambda^\bul(\frg))$. 

The monoidal structure is obvious on the classical level and given by the tensor product. There can be no quantum corrections to this result since the gauge coupling $e^2$ is an irrelevant parameter. The braiding is trivial for the same reason.

There exist yet more general line operators. To see this, we may use the dimensional reduction trick and identify the category of line operators in the 3d theory with the category of branes in the 2d theory obtained by compactifying the 3d theory on a circle. One can show that reduction  gives a B-model with target $G_\CC$ coupled to a B-type gauge theory with gauge group $G$. From the 3d viewpoint, $G_\CC$ parameterizes the holonomy of the connection $\cA$ along the compactification circle. The gauge group $G$ acts on $G_\CC$ by conjugation. As explained in appendix B, the category of branes for this TFT is the equivariant derived category of coherent sheaves $D^b_{G_\CC}(Coh(G_\CC))$. 
Line operators considered above correspond to coherent sheaves supported at the identity element of $G_\CC$. Physically, this follows from the fact that the gauge field $\cA$ is nonsingular for such line operators, and therefore the holonomy along the circle linking the line operator must be trivial. From the mathematical vewipoint, we may note that the equivariant derived category of the DG-algebra $\Lambda^\bul(\frg)$ is equivalent, by Koszul duality, to a full subcategory of the equivariant derived category of the DG-algebra 
$\Sym^\bul(\frg^*)$ (regarded as sitting in degree zero) consisting of finite-dimensional DG-modules. The latter category can also be thought as the full subcategory of the equivariant derived category of $G_\CC$ ``supported'' at the identity element of $G_\CC$. That is, focusing on line operators which are of Wilson type (i.e. are not disorder operators) is equivalent to focusing on equivariant sheaves on $G_\CC$ supported at the identity element. More generally, one may also consider Gukov-Witten-type line operators for which the conjugacy class of the holonomy of $\cA$ is fixed; such line operators can be thought of as objects of $D^b_{G_\CC}(Coh(G_\CC))$ supported at nontrivial conjugacy classes in $G_\CC$. 

\section{The gauged Rozansky-Witten model}

Recall that the Rozansky-Witten theory is a $\sigma$-model from a 3-manifold $M$ to a hyperk\"ahler target manifold $X$.  The bosonic field $\sigma$ is a map from $M$ to $X$, written locally on $X$ as $\sigma^{I}, \sigma^{\bar{I}}$ with $I = 1, 2, \cdots , \textrm{dim}_{\mathbb{C}} X$.  The fermionic fields are a 1-form on $M$ valued in $\sigma^{*}(TX)$, $\chi^{I}$, and a 0-form on $M$ valued in $\sigma^{*}(\overline{TX})$, $\eta^{\bar{I}}$.

When $X$ has a $G$-action compatible with the hyperk\"ahler structure, we can couple the corresponding Rozansky-Witten model to a B-type topological gauge theory.\footnote{For certain special $X$ and $G$, it is possible to couple the RW model to a Chern-Simons gauge theory with gauge group $G$. This gauging has been studied in \cite{KS}.}  Let $V_{a}$, $a = 1, 2, \cdots, \textrm{dim} \, G$, be the vector fields on $X$ corresponding to the generators of the $G$-action.  Let $\mu_{+}$, $\mu_{-}$, and $\mu_{3}$ be the moment maps corresponding to the holomorphic symplectic form $\Omega$, the antiholomorphic symplectic form $\bar{\Omega}$, and the K\"ahler form $J$, respectively,
\begin{align}
	d \mu_{+a} & = - i_{V_{a}} ( \Omega ) , \\
	d \mu_{-a} & = - i_{V_{a}} ( \bar{\Omega} ) , \\
	d \mu_{3a} & = i_{V_{a}} ( J ) .
\end{align}
where $i_{V}(\omega)$ is the interior product of the form $\omega$ with the vector $V$.  The BRST variation of the fields are

\begin{equation} \label{eq:SOGL_App2_GRW_variations}
	\begin{aligned}
		\delta A & = \lambda, & \delta \sigma^{I} & = 0, \\
		\delta \phi & = i \lambda, & \delta \sigma^{\bar{I}} & = \eta^{\bar{I}}, \\
		\delta \lambda & = 0, & \delta \eta^{\bar{I}} & = 0, \\
		\delta \zeta & = - i \mathcal{F}, & \delta \chi^{I} & = \mathcal{D} \sigma^{I}, \\
		\delta \rho & = i \mu_{+}, \\
		\delta \tilde{\rho} & = i P, \\ 
		\delta P & = 0.
	\end{aligned}
\end{equation}
where $\mathcal{D} \sigma^{I} = d \sigma^{I} + \mathcal{A}^{a} V_{a}^{I}$.  The action for this gauged Rozansky-Witten model is

\begin{equation}
	S = \int_{M_{3}} \big( \mathcal{L}_{1} + \mathcal{L}_{2} + \mathcal{L}_{3} + \mathcal{L}_{4} \big),
\end{equation}
with
\begin{align}
	\mathcal{L}_{1} & = - \frac{1}{2e^{2}} \delta \, \Tr \Big( i \zeta \wedge \star \bar{\mathcal{F}} + i \tilde{\rho} \wedge \star \big( P - 2 d^\star_{A} \phi - 2e^{2} \mu_{3} \big) \Big) , \\
	\mathcal{L}_{2} & = \delta \Big( g_{I \bar{J}} \chi^{I} \wedge \star \, \bar{\mathcal{D}} \sigma^{\bar{J}} \Big) , \\
	\mathcal{L}_{3} & = \delta\Big(\frac{i}{2} e^{2} \star \rho^a \mu_{-a} \Big), \\
	\mathcal{L}_{4} & = \rho_a \wedge d_{\mathcal{A}} \zeta^a  + i  \Omega_{IJ} \chi^{I} \wedge \zeta^a V^J_a + \frac{1}{2} \Omega_{IJ} \chi^{I} \wedge \mathcal{D} \chi^{J} \\
		& \quad \quad + \frac{1}{6} \Omega_{IJ} \mathcal{R}^{J}_{KL \bar{M}} \chi^{I} \wedge \chi^{K} \wedge \chi^{L} \wedge \eta^{\bar{M}} , \nonumber
\end{align}
where $\mathcal{D} \chi^{I} = d \chi^{I} + \Gamma^{I}_{JK} d \phi^{J} \wedge \chi^{K} + \mathcal{A}^{a} \partial_{J} V^{I}_{a} \wedge \chi^{J} + \mathcal{A}^{a} \Gamma^{I}_{JK} V^{K}_{a} \wedge \chi^{J}$.

Local observables in the gauged Rozansky-Witten model are BRST and gauge invariant functions of $\rho_{a}$, $\sigma^{I}$, $\sigma^{\bar{I}}$, and $\eta^{\bar{I}}$, which correspond to elements in the cohomology of $\Lambda^\bul(\mathfrak{g}) \otimes \Omega^{0, \bullet}(X)$ with respect to the following nilpotent operator,
\begin{equation}
	\delta = i \mu_{+a} T^{a} +  \bar{\partial}_{X} ,
\end{equation}
where $T^{a}$ are elements of a basis for $\mathfrak{g}$.

\section{GL-twisted theory at $t=i$ on a circle}

In this appendix we show that the GL-twisted theory at $t=i$ compactified on a circle is equivalent to a gauged version of the Rozansky-Witten model, of the sort described in the preceding appendix.   For nonabelian gauge group, the precise determination of the target space of this model is rather subtle:  naively, one can perform the compactification by simply requiring all fields to be independent of the coordinate $x^4$ of the circle, and reducing the field $A_4 + i \phi_4$ of the GL-twisted theory to a $\frg_\CC$-valued scalar $\tau$ in three dimensions.  As we will see in detail, in this case one obtains a gauged Rozansky-Witten sigma model with target $T^* \frg_\CC$, where the gauge group acts on the base by the adjoint representation and the fiber by the coadjoint representation. However, as discussed in section \ref{sec:nonabelianred}, the coordinate $\tau$ for the base is subject to global identifications, due to the possibility of performing $x^4$-dependent gauge transformations with nontrivial holonomy around the compactification circle.  Therefore, we should regard $\tau$ as merely a local coordinate on the true target space of the theory, which we conjecture to be the cotangent bundle $T^* G_\CC$. 

Let us see how the naive reduction works in detail.   
The bosonic fields in the GL-twisted theory are a 4d gauge field $A$, an adjoint-valued 1-form  $\phi$ and an adjoint-valued complex 0-form $\sigma$.  The fermionic fields are a pair of 1-forms $\psi$ and $\tpsi$, a pair of 0-forms $\eta$ and $\teta$, and a 2-form $\chi$, all adjoint-valued.

It was observed by Marcus \cite{Marcus} that, precisely at $t=\pm i$, the action can be expressed as the sum of a BRST-exact piece and a BRST-inexact fermionic piece (by contrast with the situation for $t \neq \pm i$, where the only BRST-inexact term is a purely bosonic term depending on the topology of the gauge bundle).  It is convenient to work with the complexified connections $\cA = A + i\phi$ and $\bar \cA = A - i\phi$ as well as the covariant derivatives $d_\cA$, $d_{\bar \cA}$ and curvatures $\cF$, $\bar \cF$ with respect to these connections.  We set the theta angle to zero, and place the theory on a four manifold $M_4$.  The action for GL-twisted theory at $t=i$ reads
\[
  S = \int_{M_4}  (\cL_1 + \cL_2)
\]
where
\begin{align*}
  \cL_1 &= - \frac{1}{2 e^2}\;  \delta \; \Tr \Big\{ (\chi^+ - i\chi^-) \wedge * \bar \cF   
         + d_{\bar \cA} \bsigma \wedge * (\psi - i\tpsi) \\
        &\qquad \qquad \quad + \frac{i}{2} (\eta - i\teta) \wedge *
        \Big(i[\bsigma, \sigma] - d^*_A \phi \Big) \Big\}, \\ 
  \cL_2 &= \frac{i}{e^2} \; \Tr \Big\{ (\chi^+ - i\chi^-) \wedge
        \Big( d_\cA (\psi - i\tpsi) - [\chi^+ - i\chi^- , \sigma] \Big) \Big\} 
\end{align*}
Here, * is the 4d Hodge star and $d^*_A \phi = * d_A * \phi$.     

We now take $M_4$ to be the product manifold $M_3 \times S^1$ with product metric, where $M_3$ is a three manifold and the coordinate $x^4$ ranges from $0$ to $2 \pi$ (the circumference of the $S^1$). We require fields to be independent of $x^4$, thereby obtaining an effective 3d theory on $M_3$.  It is useful to label the fields of this 3d theory as follows:
\[
	\begin{aligned}
		A^{(3d)} &= A|_{M_3},           &\sigma^{(3d)} & = \sqrt{2} \sigma,  \\
		\phi^{(3d)} &= \phi|_{M_3},     &\tau   &= (A_4 + i \phi_4 ), \\
		\cA^{(3d)} &= (A + i\phi) |_{M_3} & \eta^{\bsigma} &= \sqrt{2} i\left(\eta + i\teta \right), \\
		\bar \cA^{(3d)} &= (A - i\phi)|_{M_3} & \eta^{\btau} &= 2 i \left(\psi_4 + i \tpsi_4 \right), \\
		\lambda &= i \left(\psi + i \tpsi \right)|_{M_3}, & \chi^\sigma &= \frac{1}{\sqrt{2}} \left(\psi - i \tpsi \right)|_{M_3}, \\
		\zeta &= -i \left( \chi^+ - i\chi^- \right)|_{M_3},  &  \star \chi^\tau &= * \left(\chi^+ - i \chi^- \right)|_{M_3}, \\
		\rho &= \frac{1}{\sqrt{2}} \left( \psi_4 - i\tpsi_4 \right), & \\
		\trho &= \frac{1}{2}\left(\eta - i\teta \right).   & \\ 
	\end{aligned}  
\]
Henceforth, we drop the superscripts $(3d)$ and take $d_\cA$, etc. to refer to covariant derivatives with respect to these 3d fields.  We have written $\star$ for the 3d Hodge star and $d^\star_A \phi = \star d_A \star \phi$.  In summary, we have the following bosons: a 3d gauge field $A$, a 1-form $\phi$, and a pair of complex 0-forms $\sigma$ and $\tau$, all adjoint-valued.  We have the following fermions: a pair of 1-forms $\chi^\tau$ and  $\chi^\sigma$, a 1-form $\lambda$, a pair of 0-forms $\eta^\btau$ and $\eta^\bsigma$, another pair of 0-forms $\rho$ and $\trho$, and a 2-form $\zeta$, all adjoint-valued.  
  
In addition, it is useful to introduce an auxiliary, adjoint-valued 0-form $P$ in order to make the BRST variations nilpotent off-shell; $P$-dependent terms in the action are chosen to ensure that its equation of motion is 
\[
  P = d^\star_A \phi - \frac{i}{2}([\bsigma, \sigma] + [\btau, \tau])
\]
The dimensional reduction of the BRST variations are as follows
\[
	\begin{aligned}
		\delta A &= \lambda,          &\delta \sigma & = 0,  \\
		\delta \phi &= i \lambda,     &\delta \tau & = 0, \\ 
		\delta \lambda &= 0,          &\delta \bsigma &= \eta^{\bsigma} \\
		\delta \zeta &= - i \cF,      &\delta \btau &= \eta^{\btau} \\
		\delta \rho &= [\tau, \sigma],&\delta \eta^{\bsigma} &= 0, \\
		\delta \tilde{\rho} &= i P,   &\delta \eta^{\btau} &= 0, \\ 
		\delta P &= 0,                &\delta \chi^{\sigma} &= d_\cA \sigma, \\
		         &                    &\delta \chi^{\tau} &= d_\cA \tau.
	\end{aligned}  
\]
We have $\delta^2 = 0$ identically, without need to resort to a gauge transformation.  After an overall rescaling, the dimensional reduction of the action is as follows
\[
  S = \int_{M_3} (\cL_1 + \cL_2 + \cL_3 + \cL_4)
\]
where
\begin{align*}
  \cL_1 &= -\frac{1}{2 e^2}\;  \delta \; \Tr \Big\{i \zeta \wedge \star \bar \cF 
                   +  i \trho \wedge \star \big( P - 2 d^\star_A\, \phi 
                   + i[\bsigma, \sigma] + i[\btau, \tau] \big)  \Big\} \\
  \cL_2 &= -\frac{1}{2 e^2}\;  \delta \; \Tr \Big\{ \chi^\tau \wedge \star d_{\bar \cA} \btau
                   + \chi^\sigma \wedge  \star d_{\bar \cA} \bsigma \Big\} \\
  \cL_3 &= -\frac{1}{2 e^2}\;  \delta \; \Tr \Big\{ \rho \wedge \star [\btau, \bsigma]  \Big\}, \\                
  \cL_4 &= \frac{\sqrt{2}}{e^2} \; \Tr \Big\{ i \chi^\tau \wedge d_\cA \chi^\sigma 
        - \zeta \wedge \Big(\d_\cA\, \rho + [\chi^\sigma, \tau] 
        - [\chi^\tau, \sigma]\Big) \Big\} 
\end{align*}
The BRST-inexact piece $\cL_4$ is metric-independent, as befits a topological field theory.   We have the correct field content for a gauged Rozansky-Witten sigma model. The target space is parameterized by $\frg_\CC$-valued scalars $\sigma$ and $\tau$, both of which are acted on by the gauge group in the adjoint representation. Since $\sigma$ has ghost number $2$, we may identify the target space with $T^*[2]\frg_\CC$, where $[2]$ indicates that the fiber coordinate sits in cohomological degree $2$ (here, we are using the negative-definite quadratic form $\Tr$ to coordinatize the fiber $\frg^*_\CC$ by a $\frg_\CC$-valued scalar).  The $G$-invariant symplectic form on the target space can be read off the term $\cL_4$ of the action and is proportional to
$$
\Omega=\Tr\, d\sigma d\tau.
$$
Additionally, the $G$-invariant K\"ahler form on the target space can be inferred the term $\cL_2$ of the action and is proportional to
$$
  J = \frac{i}{2} \Tr ( d\sigma d\bsigma + d\tau d\btau )
$$
The moment maps $\mu_3, \mu_+, \mu_-$ of the $G$-action with respect to symplectic forms $J$, $\Omega$, and $\bar \Omega$ are proportional to the following quadratic functions of the coordinates: 
\begin{align*}
  \mu_3 &= -\frac{i}{2} ([\bsigma, \sigma] + [\btau, \tau]) \\
  \mu_+ &= i [\sigma, \tau] \\
  \mu_- &= - i [\bsigma, \btau]
\end{align*}
After rescaling the fields $\rho$, $\sigma$, $\tau$, $\eta^\bsigma$, $\eta^\btau$, $\chi^\sigma$, and $\chi^\tau$ by factors of $e^2$, and adjusting the relative normalization of the BRST-inexact and BRST-exact terms (which normalization does not affect the properties of the theory), one finds that the action and variations above reproduce those of a gauged Rozansky-Witten model written in the preceding appendix.

\section{A-type topological gauge theory in 2d}

In this appendix we discuss a topological gauge theory  in 2d which can be obtained by twisting $N=(2,2)$ supersymmetric gauge theory by means of the $U(1)_V$ current \cite{Witten:Agaugetheory}. This theory is somewhat analogous to the Donaldson-Witten theory in 4d, but is much simpler, because its path-integral does not get nonperturbative contributions.

The bosonic fields of the theory are a gauge field $A$ and a complex scalar $\sigma$ in the adjoint representation of the gauge group $G$. The fermionic fields are a 1-form 
$\lambda$, a 0-form $\beta$ and a 2-form $\chi$, all in the adjoint representation of $G$. The BRST transformations are
\begin{align*}
\delta A &=\lambda,\\
\delta \sigma &=0,\\
\delta\bsigma &=\eta,\\
\delta \beta &= 2i [\bsigma,\sigma],\\
\delta \lambda &=2i d_A\sigma,\\
\delta \chi &=F.
\end{align*}
The relation $\delta^2=2i\delta_g(\sigma)$ holds off-shell for all fields except $\chi$, where one has instead
$$
\delta^2\chi =d_A\lambda.
$$
If there is a fermionic equation of motion 
$$
d_A\lambda=2i[\chi,\sigma],
$$
the relation $\delta^2=2i\delta_g(\sigma)$ holds on-shell. To achieve closure off-shell we introduce an auxiliary bosonic 2-form $P$ in the adjoint representation and redefine the transformation law for $\chi$:
$$
\delta \chi =P,\quad \delta P=2i[\chi,\sigma].
$$
The action is constructed to ensure that on-shell $P=F$. One can take a BRST-exact action:
$$
S=\frac{1}{2e^2}\, \delta \int_{M_2} \left(\frac{i}{2}\lambda \wedge \stard d_A\bsigma+\chi\wedge\star (P-2F)\right)
$$
The gauge coupling enters only as the coefficient of a BRST-exact term and therefore is irrelevant. The path-integral of the theory localizes on configurations with $F=0$ and constant $\sigma$, so there is no room for nonperturbative contributions.

For purposes of this paper, we need to understand the category of branes of this 2d TFT, at least in the abelian case. We will now argue that for $G=U(1)$ the theory is isomorphic to the B-model with target $\CC[2]$, and therefore the category of branes is equivalent to $D^b(\CC[2])$.

One way to approach the problem is to construct an embedding of $D^b(Coh(\CC[2]))$ into the category of branes. This does not prove that the two categories are equivalent, but it does show that the former is a full subcategory of the latter. The basic boundary condition in the A-type 2d gauge theory is the Neumann condition which leaves the restriction of $A$ and $\sigma$ to the boundary free and requires $\stard F$ and the normal derivative of $\sigma$ to vanish on the boundary. BRST-invariance also requires $\stard\chi$ and the restriction of $\stard\lambda$ to vanish on the boundary, while $\beta$ and the restriction of $\lambda$ remain unconstrained. The algebra of BRST-invariant observables on the Neumann boundary is spanned by powers of $\sigma$, i.e. it is the algebra $\cO$ of holomorphic functions on $\CC[2]$. One can construct a more general boundary condition by placing additional degrees of freedom on the boundary which live in a graded vector space $V$. The BRST operator gives rise to a degree-1 differential $T:V\ra V$ which may depend polynomially on $\sigma$. Thus we may attach a brane to any free DG-module $M=(V\otimes \cO,T)$ over the graded algebra $\cO$. The space of morphisms between any two such branes $M_1=(V_1\otimes\cO,T_1)$ and $M_2=(V_2\otimes\cO,T_2)$ is the cohomology of the complex ${\rm Hom}_\cO(M_1,M_2)$, which agrees with the space of morphisms in the category $D^b(Coh(\CC[2]))$. 

In the A-type 2d gauge theory one may also consider the Dirichlet boundary condition which sets $\sigma=0$ on the boundary and requires the restriction of the gauge field to be trivial. One might guess that it corresponds to the skyscraper sheaf at the origin of $\CC[2]$, and indeed one can verify that the space of morphisms from any of the branes considered above to the Dirichlet brane agrees with the space of morphisms from the corresponding complex of vector bundles on $\CC[2]$ to the skyscraper sheaf. We leave this verification an an exercise for the reader.

Another way to approach the problem is construct an isomorphism between the A-type 2d gauge theory and the B-model with target $\CC[2]$.
From the physical viewpoint, an isomorphism of two 2d TFTs $\X$ and $\Y$  is an invertible topological  defect line $\A$ between them. In the present case, there is a unique candidate for such a defect line. Recall that a B-model with target $\CC[2]$ has a bosonic scalar $\phi$, a fermionic 1-form $\rho$, and fermionic 0-forms $\theta$ and $\xi$. The BRST transformations read
\begin{align*}
\delta\phi &=0,\\
\delta\bphi &=\eta,\\
\delta\eta &=0,\\
\delta \theta &=0,\\
\delta\rho &= d\phi.
\end{align*}
The field $\sigma$ has ghost number $2$, the fields $\rho$ has ghost number $1$, and the fields $\eta$ and $\theta$ have ghost number $-1$.
The action of the B-model is
$$
S=-\frac12 \delta \int_{M_2} \rho\wedge \stard d\bphi+\int_{M_2} \theta\wedge d\rho
$$
Obviously, the ghost-number $2$ bosons $\sigma$ and $\phi$ must be identified on the defect line, up to a numerical factor which can be read of the action. Similarly, the fermionic 1-forms $\lambda$ and $\rho$ must be identified, as well as the fermionic 0-forms $\beta$ and $\eta$. Finally, one must identify $\stard\chi$ and $\theta$. BRST invariance then requires $\stard F$ to vanish on the boundary, which means that the gauge field obeys the Neumann boundary condition. 

Note that this defect line is essentially the trivial defect line for the fermionic fields and $\sigma$. Since the zero-energy sector of the bosonic $U(1)$ gauge theory is trivial, the 
invertibility of the defect line is almost obvious. Let us show this more formally. First, consider two parallel defect lines with a sliver of the A-type 2d gauge theory between them. The sliver has the shape $\RR\times I$, where $\RR$ parameterizes the direction along the defect lines. The statement that the product of two defect lines is the trivial defect line in the B-model is equivalent to the statement that the $U(1)$ gauge theory on an interval with Neumann boundary conditions on both ends has a unique ground state. This is obviously true, because the space-like component of $A$ in the sliver can be gauged away by a time-independent gauge transformation, and therefore the physical phase space of the $U(1)$ gauge theory on an interval is a point.

Second, consider the opposite situation where a sliver of the B-model is sandwiched between two defect lines. We would like to show that this is equivalent to the trivial defect line in the A-type 2d gauge theory. The sliver has the shape $S^1\times I$. For simplicity we will assume that the worldsheet with a sliver removed consists of two connected components. Each component is an oriented manifold with a boundary isomorphic to $S^1$, and the path-integral of the A-type 2d gauge theory defines a vector in the Hilbert space $V$ corresponding to $S^1$. Any topological defect line in the A-type gauge theory defines an element in $V^*\otimes V\simeq {\rm End}(V)$. We would like to show that the B-model sliver corresponds to the identity element in $V^*\otimes V$. First we note that $V$ can be identified with the tensor product of the Hilbert space of the zero-energy gauge degrees of freedom and the Hilbert space of the zero-energy degrees of freedom of $\sigma$ and the fermions. As mentioned above, the defect line separating the A-type 2d gauge theory and the B-model acts as the trivial defect line on $\sigma$ and the fermions, so in this sector the statement is obvious. As for the gauge sector, the corresponding Hilbert space of zero-energy states is one-dimensional, so the B-model sliver is proportional to the identity operator. The argument of the preceding paragraph shows that its trace is one, so the sliver must be the identity operator.

The A-type 2d gauge theory with gauge group $G$ can be coupled to an A-model with target $X$ admitting a symplectic $G$-action. It is natural conjecture that the corresponding category of branes is some sort of $G$-equivariant version of the Fukaya-Floer category of $X$. To make this conjecture more precise, let us consider a special case where $X=T^*Y$ with its canonical symplectic form. It was shown in \cite{ZN,Nadler} that a suitable version of the Fukaya-Floer category of $T^*Y$ is equivalent to the constructible derived category of sheaves over $Y$. Now let $Y$ admit a $G$-action. We conjecture that the category of branes for the A-type 2d gauge theory coupled to an A-model with target $T^*Y$ is equivalent to the $G$-equivariant constructible derived category of sheaves over $Y$. In the special case when $Y$ is a point and $G=U(1)$, the latter category is known to be equivalent to $D^b(Coh(\CC[2]))$, in agreement with the fact that the A-type 2d gauge theory with $G=U(1)$ is isomorphic to the B-model with target $\CC[2]$.


\begin{thebibliography}{99}

\bibitem{Bergman} A.~Bergman, ``New Boundaries for the B-Model,'' arXiv:0808.0168 [hep-th].

\bibitem{BL} J.~Bernstein and V.~Lunts, ``Equivariant sheaves and functors,'' Lecture Notes in Math. 1578, Springer, 1994. 

\bibitem{BT} M.~Blau and G.~Thompson, ``Aspects of $N_{T}\geq 2$ Topological Gauge Theories and D-Branes,'' Nucl.\ Phys.\  B {\bf 492}, 545 (1997) [arXiv:hep-th/9612143].

\bibitem{Block}  J.~Block, ``Duality and equivalence of module categories in noncommutative geometry I,'' arXiv:math/0509284

\bibitem{LGdefects2}  I.~Brunner, H.~Jockers and D.~Roggenkamp, ``Defects and D-Brane Monodromies,''  arXiv:0806.4734 [hep-th].

\bibitem{LGdefects1} I.~Brunner and D.~Roggenkamp, ``B-type defects in Landau-Ginzburg models,''  JHEP {\bf 0708}, 093 (2007)  [arXiv:0707.0922 [hep-th]].

\bibitem{Ginz} V.~Ginzburg, ``Perverse sheaves on a loop group and Langlands duality,'' arXiv:alg-geom/9511007.

\bibitem{GNO} P.~Goddard, J.~Nuyts and D.~I.~Olive, ``Gauge Theories And Magnetic Charge,'' Nucl.\ Phys.\  B {\bf 125}, 1 (1977).

\bibitem{GW} S.~Gukov and E.~Witten, ``Gauge theory, ramification, and the geometric Langlands program,'' {\it Current developments in mathematics, 2006}, p. 35-180, Int. Press, 2008 [arXiv:hep-th/0612073]. 

\bibitem{KapVoe} M.~Kapranov and V.~Voevodsky, ``2-categories and Zamolodchikov tetrahedra equations,'' in: {\it Algebraic groups and their generalizations: quantum and infinite-dimensional methods},  177-259, Proc. Sympos. Pure Math, {\bf 56}, Part 2. American Mathematical Society, 1994.

\bibitem{Kap:qGL} A.~Kapustin, ``A Note on Quantum Geometric Langlands Duality, Gauge Theory, and Quantization of the Moduli Space of Flat Connections,''
  arXiv:0811.3264 [hep-th].

\bibitem{KR} A.~Kapustin and L.~Rozansky, ``Three-dimensional topological field theory and symplectic algebraic geometry II,''  arXiv:0909.3643 [math.AG].

\bibitem{KRS} A.~Kapustin, L.~Rozansky and N.~Saulina, ``Three-dimensional topological field theory and symplectic algebraic geometry I,''
 Nucl.\ Phys.\  B {\bf 816}, 295 (2009) [arXiv:0810.5415 [hep-th]].
 
 \bibitem{KS} A.~Kapustin and N.~Saulina, ``Chern-Simons-Rozansky-Witten topological field theory,'' Nucl.\ Phys.\  B {\bf 823}, 403 (2009)  [arXiv:0904.1447 [hep-th]].

\bibitem{KW} A.~Kapustin and E.~Witten, ``Electric-magnetic duality and the geometric Langlands program,''  Commun. Number Theory Phys. {\bf 1}, 1 (2007).

\bibitem{Lusztig}  G.~Lusztig, ``Singularities, character formula, and a q-analog of weight multiplicities,'' Analyse et Topologie Sur Les Espaces Singuliers II-III, Ast\'{e}rique {\bf 101-102}, 208 (1983).

\bibitem{Marcus} N.~Marcus, ``The Other topological twisting of N=4 Yang-Mills,'' Nucl.\ Phys.\  B {\bf 452}, 331 (1995)  [arXiv:hep-th/9506002].

\bibitem{MV} I.~Mirkovic and K.~Vilonen, ``Geometric Langlands duality and representations of algebraic groups over commutative rings,'' Ann. of Math. (2) {\bf 166}, 95 (2007).


\bibitem{Nadler} D.~Nadler, ``Microlocal branes and constructible sheaves,'' arXiv:math/0612399.

\bibitem{ZN} D.~Nadler and E.~Zaslow, ``Constructible sheaves and the Fukaya category,'' J. Amer. Math. Soc. {\bf 22}, 233 (2009) [arXiv:math/0604379]

\bibitem{ToVe} B.~Toen and G.~Vezzosi, ``A note on Chern character, loop spaces and derived algebraic geometry'', arXiv:0804.1274 [math.AG].

\bibitem{Witten:Donaldson} E.~Witten, ``Topological Quantum Field Theory,'' Commun.\ Math.\ Phys.\  {\bf 117}, 353 (1988).

\bibitem{Witten:Agaugetheory} E. Witten, in: ``Quantum Fields and Strings: a course for mathematicians,'' section 14.4, vol. 2, ed. by P.~Deligne et al, AMS, 1999.

\bibitem{Witten:wild}  E.~Witten, ``Gauge Theory And Wild Ramification,'' arXiv:0710.0631 [hep-th].









\end{thebibliography}
\end{document}